\documentclass[%
 aip, amsmath,amssymb, reprint, floatfix
]{revtex4-1}

\usepackage{graphicx}
\usepackage{dcolumn}
\usepackage{bm}

\usepackage[utf8]{inputenc}
\usepackage[T1]{fontenc}
\usepackage{mathptmx}
\usepackage{etoolbox}

\makeatletter
\def\@email#1#2{%
 \endgroup
 \patchcmd{\titleblock@produce}
  {\frontmatter@RRAPformat}
  {\frontmatter@RRAPformat{\produce@RRAP{*#1\href{mailto:#2}{#2}}}\frontmatter@RRAPformat}
  {}{}
}%
\makeatother

\usepackage{empheq}
\draft 

\usepackage{bm}
\usepackage{xcolor}
\usepackage{soul}

\usepackage{tikz}
\usetikzlibrary{positioning}
\usetikzlibrary {arrows.meta}

\newcommand{\hx}{\widehat{x}}
\newcommand{\hz}{\widehat{z}}
\newcommand{\hP}{\widehat{P}}
\newcommand{\hv}{\widehat{v}}

\newcommand{\bP}{\overline{P}}
\newcommand{\bv}{\overline{v}}

\newcommand{\derhx}[1]{\frac{\partial{#1}}{\partial \widehat{x}}}
\newcommand{\derhz}[1]{\frac{\partial{#1}}{\partial \widehat{z}}}
\newcommand{\derderhx}[1]{\frac{\partial^2{#1}}{\partial^2 \widehat{x}}}
\newcommand{\derderhz}[1]{\frac{\partial^2{#1}}{\partial^2 \widehat{z}}}

\usepackage{bigints}

\usepackage{tikz}
\DeclareRobustCommand\full  {\tikz[baseline=-0.6ex]\draw[thick] (0,0)--(0.5,0);}
\DeclareRobustCommand\dotted{\tikz[baseline=-0.6ex]\draw[thick,dotted] (0,0)--(0.54,0);}
\DeclareRobustCommand\dashed{\tikz[baseline=-0.6ex]\draw[thick,dashed] (0,0)--(0.54,0);}

\usepackage{mathtools}
\usepackage{upgreek}
\usepackage{mhchem}

\begin{document}

\title{The interplay of shape and catalyst distribution in the yield of compressible flow microreactors}

\author{G. C. Antunes}
\affiliation{Helmholtz-Institut Erlangen-Nürnberg für Erneuerbare Energien (IET--2), Forschungszentrum J\"ulich, Cauerstr.~1, 91058 Erlangen,\,Germany}

\author{M. Jim\'enez-S\'anchez}
\affiliation{Department of Chemistry and Pharmacy, Chair: Chemistry of Thin Film Materials, Friedrich-Alexander-Universit\"at Erlangen-N\"urnberg (FAU), 91058 Erlangen, Germany}

\author{P. Malgaretti}
 \affiliation{Helmholtz-Institut Erlangen-Nürnberg für Erneuerbare Energien (IET--2), Forschungszentrum J\"ulich, Cauerstr.~1, 91058 Erlangen,\,Germany}
 \email{p.malgaretti@fz-juelich.de}

\author{J. Bachmann}
\affiliation{Department of Chemistry and Pharmacy, Chair: Chemistry of Thin Film Materials, Friedrich-Alexander-Universit\"at Erlangen-N\"urnberg (FAU), 91058 Erlangen, Germany}
 
 \author{J. Harting}
\affiliation{Helmholtz-Institut Erlangen-Nürnberg für Erneuerbare Energien (IET--2), Forschungszentrum J\"ulich, Cauerstr.~1, 91058 Erlangen,\,Germany}
\affiliation{Department Chemie- und Bioingenieurwesen und Department Physik, Friedrich-Alexander-Universit\"at Erlangen-N\"urnberg, F\"{u}rther Stra{\ss}e 248, 90429 N\"{u}rnberg, Germany}


\begin{abstract}
We develop a semi-analytical model for transport in structured catalytic microreactors, where both reactant and product are compressible fluids. Making use of the lubrication and Fick-Jacobs approximations, we reduce the three-dimensional governing equations to an effective one-dimensional set of equations. Our model captures the effect of compressibility, of corrugations in the shape of the reactor, as well as of an inhomogeneous catalytic coating of the reactor walls. We show that in the weakly compressible limit (e.g., liquid-phase reactors), the distribution of catalyst does not influence the reactor yield, which we verify experimentally. Beyond this limit, we show that introducing inhomogeneities in the catalytic coating and corrugations to the reactor walls can improve the yield. 

\end{abstract}

\maketitle

\section{Introduction}

Catalysts and catalytic reactors are ubiquitous in the modern-day world. They are used in the production of organic and inorganic chemicals \cite{book_IndustrialCatalysis}, in crude oil refining \cite{Tanimu2022}, in energy conversion \cite{Zhu2020, Pei2018, book_FuelCells}, and for environmental protection / green chemistry \cite{Anastas2001}. The latter has become increasingly more relevant as the world deals with the consequences of climate change \cite{Anastas2002, Zimmerman2020}. Research into novel catalysts is ever ongoing, with great attention being given to recent developments such as, e.g., supported ionic liquid phase (SILP) catalysis \cite{Riisager2006,Riisager2008}, supported catalytically active liquid metal solutions (SCALMS)\cite{Taccardi2017,Rupprechter2017}, and solid catalysts with ionic liquid layer (SCILL)\cite{Kernchen2007}. Catalysts such as these are deposited within a reactor through which fluid containing reactant is flowed. These reactors may be slabs through which channels run through \cite{Boger2004}, or compressed solid grains/particles \cite{Onsen2016}. Common models for the transport through such media rely on coarse-graining the porous structure. These models do not consider the pores explicitly, instead treating the flow through the porous media as flow in free space with effective transport parameters \cite{Woudberg2008,Macdonald1979}. Such models often follow from averaging the flow quantities over a volume much larger than the typical pore size \cite{Bachmat1986, Bear1986}, and as such, fail when examining flow at the pore scale \cite{Negrini1999,Battiato2011}. Furthermore, 

A well-known use of catalysts is in the automotive industry. The functioning of motor vehicles leads to the production of environmentally dangerous chemicals, which are eliminated by catalysts rather than emitted into the atmosphere \cite{book_catalyticAirPollution}. A common reactor design is that of the so-called monolithic reactor, composed of a slab of material through which channels run through \cite{Boger2004}. The walls of these channels are coated with catalytic material. Catalytic porous media such as packed-bed reactors are also an industry-standard \cite{Onsen2016}. Common models for the transport through such media rely on coarse-graining the porous structure. These models do not consider the pores explicitly, but instead treat the flow through the porous media as flow in free space with effective transport parameters \cite{Woudberg2008,Macdonald1979}. Such models often follow from averaging the flow quantities over a volume much larger than the typical pore size \cite{Bachmat1986, Bear1986}, and as such, fail when examining flow at the pore scale \cite{Negrini1999,Battiato2011}. 

Understanding the transport of reacting gases at the microscale is also vital for the development and study of miniaturized reactors, which have received great attention in the last decades \cite{Ran2023,Kolb2004,DeWitt1999}. Research and development of new chemical processes benefits from the decreased amount of sample required in microfluidic devices, as well as from the heightened control over the chemical conditions inside the microreactor coming from higher surface-to-volume ratios \cite{Ran2023,Kolb2004,DeWitt1999}. Besides an increase in efficiency and decrease in cost \cite{DeWitt1999}, conducting research in miniaturized reactors leads to a decrease in the production of environmentally unsafe chemicals \cite{Anastas2010}. Indeed, an effort to make the chemical industry greener has taken root in the last decades, and is expected to play a vital role in the mitigation of climate change \cite{Anastas2010}. Large-scale production of chemicals also benefits from employing multiple microreactors in parallel rather than a single, large batch reactor. The increased control over the chemical properties inside the reactor may facilitate operation at very efficient conditions that are not easily enforceable in batch reactors \cite{Ran2023,Kolb2004,DeWitt1999}. Furthermore, the increased control leads to the entire process becoming safer, as the chance of a malfunction decreases \cite{DeWitt1999}.

Here, we present a theory for the flow of a compressible fluid in a quasi-2D reactor/pore whose walls are coated with catalyst. To obtain analytical results, we make use of the Fick-Jacobs approximation, which is valid for thin reactors. As the reactant flows through the reactor, it turns into the product in the presence of the catalyst. The theory enables the study of reactors with corrugated walls and a spatially varying reaction rate. As a result, we are able to optimize reactor parameters such as the shape of the corrugation, and the distribution of catalytic material. We predict that the distribution of catalytic material has no effect on the flow rate of the product in the incompressible limit and verify this result experimentally.

\section{The model}

\begin{figure}[t]
	\centering
   \includegraphics[width=0.5\textwidth]{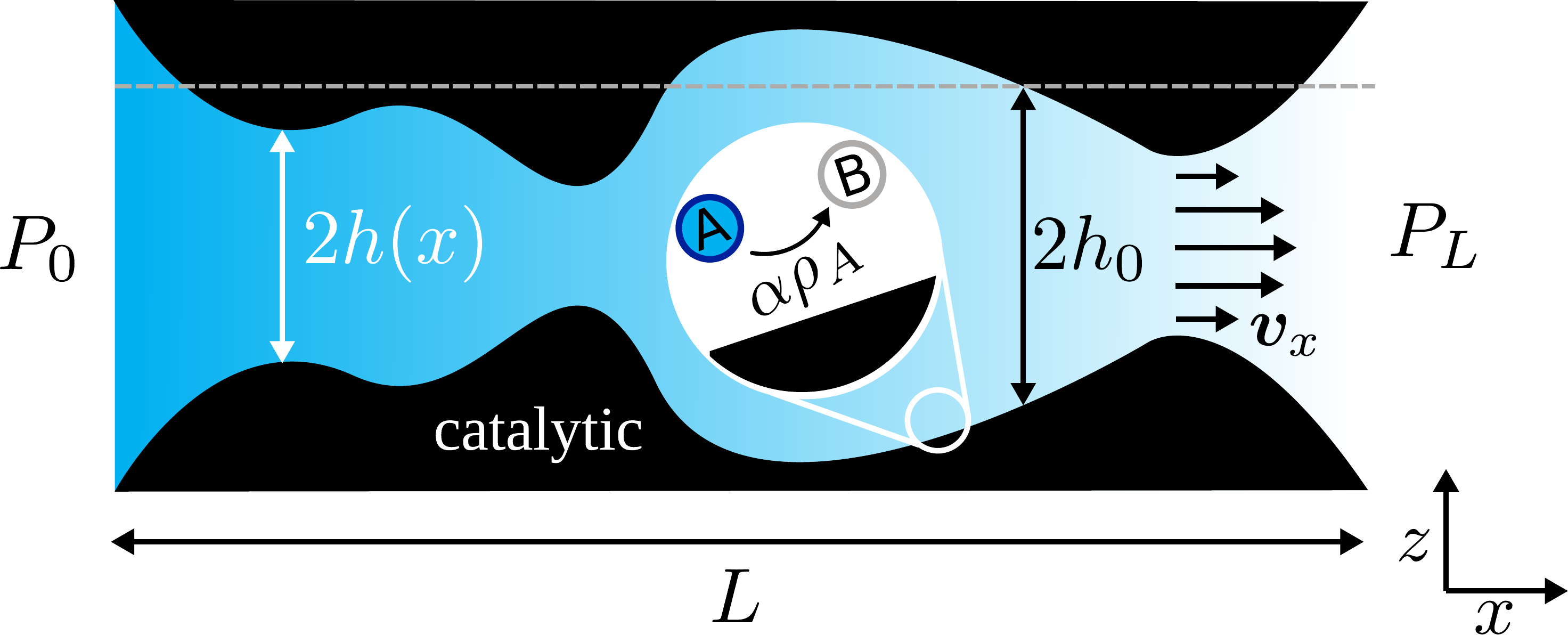}
   \caption{ Longitudinal section of the quasi-2D catalytic reactor with length $L$ and variable radius $h(x)$, the average of which is $h_0$. The irreversible decomposition of a chemical species $A$ to produce a chemical species $B$ occurs solely at the inner reactor wall with rate $\alpha \rho_A$. The color blue indicates the value of the total number density and, therefore, of the pressure. The darker the blue, the higher the pressure. The pressure at inlet $P_0$ is different from the pressure at outlet $P_0$, leading to a flow in the $x$ direction $v_x$. \label{fig:Paper3_model}}
\end{figure}

Consider a reactor defined by the space between two solid slabs. These slabs are symmetric with respect to the axis $z=0$, and are homogeneous in the $y$ direction. The reactor is thus quasi-3D. The slabs are corrugated in the $x$ direction, with the separation between them defined as $2h(x)$, as in Fig.~\ref{fig:Paper3_model}. It will be useful to define the average channel height
\begin{equation}
    h_0 = \frac{1}{L} \int\limits^L_0 h(x) dx.
\end{equation}
The length of the reactor is $L$, such that to the left ($x<0$) and to the right ($x>L$) of the slabs are two reservoirs of gas with pressure $P_0$ and $P_L$, respectively. The gas fills the reactor as well, with a spatially varying pressure $P(x,z,t)$, such that
\begin{equation}
    P(0,z,t) = P_0
\end{equation}
and
\begin{equation}
    P(L,z,t) = P_L.
\end{equation}
Due to the translational symmetry in the $y$ direction, the pressure depends only on the $x$ and $z$ direction. The pressure is related to the \textit{number} density of the gas via the virial expansion
\begin{equation}
    P(x,z,t) = k_BT \rho(x,z,t)\left[ 1 + B_1\rho(x,z,t) + B_2\rho^2(x,z,t) + ... \right], \label{eq:Paper3_virialGas}
\end{equation}
where $k_B$ is the Boltzmann constant, $T$ is the absolute temperature, $\rho(x,z,t)$ is the \textit{number} density of the gas (with dimensions $\text{meter}^{-3}$, and $B_i \ (i\geq 1)$ are the virial coefficients. Many gasses (e.g. methane, water vapor, oxygen, nitrogen) exhibit values of $|B_1|$ below $\approx 10^2  \textrm{cm}^3/\textrm{mol}$ for $T\approx 400-1000 \textrm{K}$ \cite{Sengers1971}. Typically, gasses (including the four aforementioned examples) in that temperature range and for pressures around 1 bar exhibit densities $\approx 10^{-5} \textrm{mol}/\textrm{cm}^3$ \cite{NIST}. These densities are well below $B_1^{-1} \approx 10^{-2} \textrm{mol}/\textrm{cm}^3$, and thus the ideal gas equation of state is a good approximation
\begin{equation}
    P(x,z,t) \approx  k_BT \rho(x,z,t). \label{eq:Paper3_idealGas}
\end{equation}
The pressure difference between the ends of the reactor leads to a non-zero flow velocity $\bm{v}(x,z,t)$ inside the reactor as the gas moves from one reservoir to the other. The flow velocity is governed by the (compressible) Navier-Stokes equation
\begin{align}
&\rho_f(x,z,t) \left\{\dot{\bm{v}}(x,z,t) + [\bm{v}(x,z,t) \cdot \nabla]  \bm{v}(x,z,t) \right\} = \nonumber \\
&= - \nabla P(x,z,t) + \eta \Delta \bm{v}(x,z,t) + \left (\zeta + \frac{1}{3}\eta \right) \nabla [\nabla \cdot \bm{v}(x,z,t)],
\label{eq:Paper3_NS} 
\end{align}
where $\rho_f(x,z,t)$ is the \textit{mass} density of fluid, and $\eta$ and $\zeta$ are the shear and bulk viscosity, respectively. No bulk force is applied to the fluid. The slabs, the reactor, and the reservoirs are kept at a constant temperature, and thus the viscosities $\eta$ and $\zeta$ are homogeneous in space and constant in time as well~\cite{LeBellac2004, Kumar2019}. Equation \eqref{eq:Paper3_NS} is subjected to no-slip boundary conditions on the reactor walls
\begin{equation}
    \bm{v}[x,\pm h(x), t] = 0. \label{eq:noslip}
\end{equation}
Furthermore, the continuity equation applies
\begin{align}
\dot{\rho_f}(x,z,t) & =-\nabla\cdot\left[\rho_f(x,z,t) \bm{v}(x,z,t) \right]. \label{eq:Paper3_continuity} 
\end{align}
The walls of the reactor are coated with a catalytic material that decomposes a reactant gas coming from the inlet ($A$ particles) and converts it into a product gas ($B$ particles). For the remainder of this work, $A$ and $B$ particles are physically indistinguishable (same mass, same size, etc.). The chemical reaction is thus merely a change of label. Since $A$ and $B$ particles have the same mass $m$, we write
\begin{equation}
\rho_f(x,z,t) = m \rho(x,z,t) ,
\end{equation} 
and
\begin{equation}
\rho(x,z,t)  = \rho_A(x,z,t)  + \rho_B(x,z,t) , \label{eq:Paper3_defTotRho}
\end{equation} 
where $\rho_A(x,z,t) $ and $\rho_B(x,z,t) $ are the number densities of species A and B, respectively. These number densities are governed by the advection-diffusion equations
\begin{equation}
\dot{\rho_A}(x,z,t) =-\nabla\cdot\bm{j}_A(x,z,t)-\xi(x,z,t), \label{eq:Paper3_ADE_A}
\end{equation}
and
\begin{equation}
\dot{\rho_B}(x,z,t) =-\nabla\cdot\bm{j}_B(x,z,t)+\xi(x,z,t), \label{eq:Paper3_ADE_B}
\end{equation}
where $\bm{j}_A(x,z,t)$ and $\bm{j}_B(x,z,t)$ are the fluxes of species $A$ and $B$ respectively, and $\xi(x,z,t)$ is the rate of reaction. For simplicity, we assume a first-order reaction rate \cite{book_Atkins}
\begin{equation}
\xi(x,z,t) = \alpha(x) \rho_A(x,z,t)\left\{\delta[z-h(x)]+\delta[z+h(x)]\right\}, \label{eq:Paper3_xi} 
\end{equation}
where $\delta(z)$ is the Dirac delta function, which localizes the chemical reaction at the reactor walls. The function $\alpha(x)$ is a spatially varying reaction rate with unit meter/sec, as can be seen from the dimensional analysis of Eqs.~\eqref{eq:Paper3_ADE_B}  and \eqref{eq:Paper3_xi} \footnote{It is worthwhile to note that $\delta(z)$ has unit meter$^{-1}$, as can be seen from how the integral $\int\limits_{-a}^a \delta(z) dz =1$ is dimensionless (for any value of $a \neq 0$) \cite{book_MathMeth}.}.  The fact that $\alpha(x)$ is spatially varying accounts for a possibly inhomogeneous concentration of catalytic material coating the reactor wall. It will be useful to define the spatially-averaged reaction rate $\alpha_0$ such that
\begin{equation}
\alpha_0 = \frac{1}{L} \int\limits_0^L \alpha(x) dx,
\end{equation}
and 
\begin{equation}
    \alpha(x) = \alpha_0 + \delta \alpha (x), \label{eq:defalpha1}
\end{equation}
where $\delta \alpha(x)$ is the deviation from the average reaction rate.
Equations \eqref{eq:Paper3_ADE_A} and \eqref{eq:Paper3_ADE_B} are solved with the added constraint that gas cannot escape through the reactor walls
\begin{align}
\rho_A[x, |z| > h(x)] &= 0, \label{eq:Paper3_noAoutside}\\
\rho_B[x, |z| > h(x)] &= 0, \label{eq:Paper3_noBoutside}
\end{align}
and as such 
\begin{align}
\bm{j}_A[x, |z| > h(x)] &= 0, \label{eq:Paper3_nofluxAoutside}\\
\bm{j}_B[x, |z| > h(x)] &= 0.\label{eq:Paper3_nofluxBoutside}
\end{align}
The fluxes $\bm{j}_A(x,z,t)$ and $\bm{j}_B(x,z,t)$ can be decomposed into two components,
\begin{equation}
\bm{j}_A(x,z,t) = \rho_A(x,z,t) \bm{v}(x,z,t) + \bm{j}_D^{(A)}(x,z,t), \label{eq:Paper3_flowA} 
\end{equation}
and
\begin{equation}
\bm{j}_B(x,z,t) = \rho_B(x,z,t) \bm{v}(x,z,t) + \bm{j}_D^{(B)}(x,z,t), \label{eq:Paper3_flowB} 
\end{equation}
with the first term in the right-hand-side of Eqs.~\eqref{eq:Paper3_flowA} and \eqref{eq:Paper3_flowB} accounting for the motion of the fluid mixture as a whole, and the second term in the right-hand-side accounting for diffusion. The velocity $\bm{v}(x,z,t)$ is the center-of-mass/barycentric velocity of the mixture \cite{Lam2006,Krishna1993}. In addition to this motion, $A$ and $B$ particles may move relative to another one, which is captured by the diffusive fluxes $\bm{j}_D^{(A)}(x,z,t)$ and $\bm{j}_D^{(B)}(x,z,t)$ \cite{Lam2006, book_NEThermo}. Because the gas is well-approximated as ideal, the diffusive fluxes may be written as Fick's law \cite{Krishna1993},
\begin{align}
\bm{j}_D^{(A)}(x,z,t) = - D\rho (x,z,t)\nabla \left( \frac{\rho_A(x,z,t)}{\rho(x,z,t)} \right),\\
\bm{j}_D^{(B)}(x,z,t) = - D\rho (x,z,t)\nabla \left( \frac{\rho_B(x,z,t)}{\rho(x,z,t)} \right),
\end{align}
where $D$ is the diffusion coefficient, which does not vary in space or time \cite{Krishna1993}. Note that the sum of the diffusive fluxes is zero
\begin{align}
&\bm{j}_D^{(A)}(x,z,t) + \bm{j}_D^{(B)}(x,z,t) = \nonumber \\
& =- D\rho(x,z,t) \nabla \left( \frac{\rho_A(x,z,t) + \rho_B(x,z,t)}{\rho(x,z,t)} \right)= \nonumber \\
&= - D\rho(x,z,t) \nabla \left( \frac{\rho(x,z,t)}{\rho(x,z,t)} \right)  = 0,
\end{align}
as is required from the fact that the diffusive fluxes are defined as the motion of $A$ and $B$ particles with respect to the center of mass of the binary mixture \cite{Lam2006,Krishna1993,book_NEThermo}. We assume absence of product in the inlet
\begin{equation}
\rho_B (x=0,z,t) = 0,
\end{equation} 
which together with the inlet pressure $P_0$ and outlet pressure $P_L$ form the set of boundary conditions needed to solve Eqs.~\eqref{eq:Paper3_ADE_A} and \eqref{eq:Paper3_ADE_B}: 
\begin{equation}
\rho_A(x=0,z,t) = \frac{P_0}{k_BT},
\end{equation}
and 
\begin{equation}
\rho_A(x=L,z,t) + \rho_B(x=L,z,t) = \frac{P_L}{k_BT}.
\end{equation}

\section{Reduction to effective 1D equation}

Solving Eqs.~\eqref{eq:Paper3_NS}, \eqref{eq:Paper3_ADE_A}, and \eqref{eq:Paper3_ADE_B} in the general case presents significant difficulties. Instead, we specialize Eqs.~\eqref{eq:Paper3_NS}, \eqref{eq:Paper3_ADE_A}, and \eqref{eq:Paper3_ADE_B} to the case of a thin reactor ($h_0/L \ll 1$). Such a condition is often true for monolithic reactors \cite{Boger2004, Govender2017, Kapteijn2013}. The separation of length scales leads to relaxation in the transverse direction, which is much faster than relaxation in the longitudinal direction. As a result, we are able to describe the dynamics in the longitudinal direction via an effective 1D equation. This is the so-called Fick-Jacobs approximation~\cite{Zwanzig1992,Reguera2001,Kalinay2005,Kalinay2008,Martens2011,Dagdug2013,Malgaretti2023} for transport in thin channels. Such a technique has been used to model transport of colloids~\cite{Reguera2006,Reguera2012,Marconi2015,Malgaretti2016_entropy,Puertas2018}, polymers~\cite{Bianco2016,Locatelli2023}, electrolytes~\cite{Malgaretti2014,Chinappi2018,Malgaretti2019_JCP}, and chemically active systems~\cite{Dagdug2014,Kalinay2022,Antunes2022, Antunes2023}.  This procedure has the added benefit of yielding simple formulas that provide analytical insight and are computationally cheap to evaluate.

As we are interested in the long-term functioning of the reactor, our starting point is to disregard transients in the dynamics and to focus on the steady state. We will now solve for the steady-state flow of mixture through the reactor. At steady state, the continuity equation (Eq.~\eqref{eq:Paper3_continuity}) yields 
\begin{equation}
0  =-\nabla\cdot\bm{j}(x,z),  \label{eq:Paper3_contSteady}
\end{equation}
where 
\begin{equation}
\bm{j}(x,z) = \bm{j}_A(x,z) + \bm{j}_B(x,z) = \rho(x,z) \bm{v}(x,z)
\end{equation}
is the total flux of the mixture. Integrating Eq.~\eqref{eq:Paper3_contSteady} in the transverse direction leads to
\begin{equation}
0 =\int\limits_{- \infty}^{ \infty} \left[ \partial_x \bm{j}_x(x,z) + \partial_z \bm{j}_z(x,z) \right]dz, \label{eq:Paper3_contSteady1}
\end{equation}
where we take the integration limits to be $\pm \infty$ for ease of future calculations. We shall see that all results will reduce to integrals inside the reactor. Eq.~\eqref{eq:Paper3_contSteady1}  reduces to
\begin{equation}
0 = \partial_x \left( \int\limits_{-\infty}^{\infty} \bm{j}_x(x,z) dz \right), \label{eq:Paper3_constTotFlux00}
\end{equation}
as there is no flux at infinity. Indeed, there is no flux outside of the reactor (Eq.~\eqref{eq:Paper3_nofluxAoutside} and \eqref{eq:Paper3_nofluxBoutside}), which leads to
\begin{equation}
0 = \partial_x \left( \int\limits_{-h(x)}^{h(x)} \bm{j}_x(x,z) dz \right). \label{eq:Paper3_constTotFlux0}
\end{equation}
We now define the total flow $J(x)$ across the reactor as the integral of the flow density $\bm{j}_x(x,z)$, 
\begin{equation}
J(x) = \int\limits_{-h(x)}^{h(x)} \bm{j}_x (x,z)dz, \label{eq:Paper3_defJ}
\end{equation}
which inserted in Eq.~\eqref{eq:Paper3_constTotFlux0} yields
\begin{equation}
\partial_x J(x) = 0, \label{eq:Paper3_constTotFlux}
\end{equation}
as is required for a steady state. The total flux $J$ is a flux of particles per unit length, as the integral in Eq.~\eqref{eq:Paper3_contSteady1} is done over $z$ only (due to the quasi-3D geometry of the reactor). As discussed earlier, we aim to reduce the quasi-3D problem to an effective one-dimensional problem by employing the Fick-Jacobs approximation. This approximation exploits the difference in timescales characterizing transverse and longitudinal transport. 

As discussed, we assume a thin reactor
\begin{equation}
    \frac{h_0}{L} \ll 1, \label{eq:thinReactor}
\end{equation}
which allows for an approximated solution of Eq.~\eqref{eq:Paper3_NS} by employing the lubrication approximation \cite{Schlichting1979}. Such a procedure is done in detail in Appendix \ref{sec:1D}, and yields the flow velocity profile   
\begin{equation}
 v_{x}(x,z)=\frac{z^{2}-h^{2}(x)}{2\eta}\partial_x P(x)\,, \label{eq:Paper3_v_x_MAIN}
\end{equation}
provided that the channel is thin enough so that
\begin{align}
    &\frac{h_0}{L} Re \ll 1, \label {eq:condReynolds} \\
    &\left( \frac{h_0}{L}\right)^2 \frac{\zeta}{\eta} \ll 1. \label{eq:noCompress}
\end{align}
where we define the Reynolds number $Re$ as
\begin{align}
    Re   &= \bar{\rho}_f \frac{\bar{v}_x h_0}{\eta}, 
\end{align}
where $\bar{\rho}_f$ and $\bar{v}_x$ are the typical scales for mass density and longitudinal velocity, respectively. The condition in Eq. \eqref{eq:condReynolds} is obtained in thin channels ($h_0 \ll L$) provided that $Re$ is itself not a large number. While the latter generally holds for flows in micrometric channels, Eq.~\eqref{eq:condReynolds} will be seen as imposing a maximum height for the reactor (given a certain reactor aspect ratio).

The condition in Eq.~\eqref{eq:noCompress} requires the knowledge of the ratio of bulk viscosity to shear viscosity. For some gasses, the ratio $\zeta/\eta$ is smaller than 1 ( e.g., carbon monoxide), or order one (e.g., methane) \cite{Cramer2012}, and Eq.~\eqref{eq:noCompress} is fulfilled whenever Eq.~\eqref{eq:thinReactor} is as well. However, it is also possible for the ratio $\zeta/\eta$ to take values as large as $\approx 4000$, such as for carbon dioxide \cite{Cramer2012}. The conditions in Eqs.~\eqref{eq:condReynolds} and  \eqref{eq:noCompress} further imply that the variation of pressure along the transverse direction is negligible (see Appendix \ref{sec:1D}),
\begin{equation}
    P(x,z) \equiv P(x).
\end{equation}
Besides the advective contribution, the flow of product and reactant also contains a diffusive contribution as per Eqs.~\eqref{eq:Paper3_flowA}  and \eqref{eq:Paper3_flowB}. Due to the length scale separation between reactor length and height, it is possible for either diffusion or advection to dominate transport in certain directions. The competition between diffusion and advection is quantified via two P\'eclet numbers (one for each direction). For the longitudinal direction, we have
\begin{align}
    Pe_x &= \frac{\overline{v}_x L}{D},
\end{align}
and in the transversal direction
\begin{align}
    Pe_z &= \frac{\overline{v}_z h_0}{D},
\end{align}
 where $\bar{v}_z$ is the typical velocity in the $z$ direction. As a result of Eq.~\eqref{eq:thinReactor}, we have
 \begin{equation}
     \frac{\bar{v}_z}{\bar{v}_x} \approx \frac{h_0}{L}, \label{eq:condVxVy}
 \end{equation}
 which is rigorously derived in Appendix \ref{sec:1D}. To obtain an analytical result, we now specialize our theory to the case where advection is the dominant transport mechanism in the longitudinal direction, i.e.
 \begin{equation}
     Pe_x \gg 1, \label{eq:CondPex}
 \end{equation}
 and so we approximate
\begin{align}
(\bm{j}_A )_x(x,z) &\approx \rho_A(x,z) \bm{v}_x(x,z), \label{eq:Paper3_fluxA_lon_MAIN}  \\
(\bm{j}_B )_x(x,z) &\approx \rho_B(x,z) \bm{v}_x(x,z). \label{eq:Paper3_fluxB_lon_MAIN}
\end{align}
Conversely, in the transversal direction, we assume diffusion to be the dominant transport mechanism, with
\begin{equation}
    Pe_y \ll 1,  \label{eq:CondPey}
\end{equation}
which by use of Eq.~\eqref{eq:condVxVy} yields a maximum value for
\begin{equation}
    Pe_x \ll \left( \frac{L}{h_0} \right)^2. \label{eq:CondPex2}
\end{equation}
In this regime, we obtain
\begin{align}
(\bm{j}_A )_z(x,z) &\approx D \partial_z \rho_A(x,z) , \label{eq:Paper3_fluxA_lon_MAIN_z}  \\
(\bm{j}_B )_z(x,z) &\approx D \partial_z \rho_B(x,z) . \label{eq:Paper3_fluxB_lon_MAIN_z}
\end{align}
It is worth noting that Eqs.~\eqref{eq:CondPex} and \eqref{eq:CondPey} can be written as restrictions on $h_0$ and $L$. By approximating $\bv_x$ via the plane Poiseuille law (averaged over the reactor height)\cite{book_Russianguy}, one obtains
\begin{align}
        h_0 &\gg \left(  \frac{3 D \eta }{\Delta P} \right)^{1/2} , \\
        L &\gg  h_0^2\left( \frac{3 D \eta }{\Delta P} \right)^{-1/2} .
\end{align} 
Using typical values of  $\eta \simeq 5 \times 10^{-5}\,\text{Pa}\cdot\text{s}$ \cite{Nagashima1985}, $D \simeq 10^{-5}\text{m}^2/\text{s}$ \cite{Wilke1955}, and $\Delta P\simeq 10^5 \text{Pa}$, one finds $  (3 D \eta /\Delta P)^{1/2}  \approx 0.1 \mu m$. As such, we expect our theory to be valid for reactors with micrometric width and relatively large aspect ratios $L/h_0$.

Under the assumption of Eq.~\eqref{eq:CondPey}, the dependence of the concentration of reactant and product in the transversal direction is determined by diffusion and the reaction at the wall. Diffusive flows promote homogeneity in $z$, whereas the reaction promotes gradients in concentration. The competition of these two effects is quantified via the Damk\"ohler number along the transversal direction,
\begin{equation}
    Da_z = \frac{\alpha_0 h_0}{D}.
\end{equation}
In reactors that are small enough for $Da_z$ to be small, i.e.
\begin{equation}
    Da_z \ll 1, \label{eq:condDam}
\end{equation}
the concentration of both $A$ and $B$ particles may be approximated as independent of the transversal direction  (see Appendix \ref{sec:1D}):
\begin{align}
    \rho_A(x,z) &\approx \rho_A(x), \label{eq:Paper3_Aconst}\\
    \rho_B(x,z) &\approx \rho_B(x), \label{eq:Paper3_Bconst}
\end{align}
for $|z| < h(x)$. The condition in Eq.~\eqref{eq:condDam} is fulfilled whenever the average reaction rate $\alpha_0$ is such that
\begin{equation}
    \alpha_0 \ll \frac{D}{h_0}, \label{eq:alpha0Cond}
\end{equation}
which for $h_0 \approx 1 \mu m$ yields the condition $\alpha_0 \ll 10$ m/s. Note that this velocity quantifies the reaction rate and is not the velocity at which gas is flowing. Examples of environmentally and economically relevant reactions that fit this criteria are methane/air mixtures in contact with platinum  ($\alpha_0 < 10^{-1} \text{m/s}$ for temperature below $1100K$ \cite{Reinke2004}), carbon monoxide/oxygen mixtures in contact with platinum ($\alpha_0 < 1 \text{m/s}$ for temperature below $305 K$\cite{Papp2013}, and ammonia production in sustainable reactors (e.g. electrochemical cells, $\alpha_0 < 10^{-5} \text{m/s}$ \ \cite{Ghavam2021,Xu2009}).

Furthermore, note that decreasing the reaction rate of a catalyzed reaction can often be achieved by simply decreasing the amount of catalytic material coating the reactor wall. As such, we proceed with the condition in Eq.~\eqref{eq:condDam}, from which Eqs.~\eqref{eq:Paper3_Aconst} and \eqref{eq:Paper3_Bconst} result. Finally, the theory assumes a Knudsen number $Kn$ small enough to justify using the no-slip boundary condition (Eq. \eqref{eq:noslip})
\begin{equation}
Kn = \frac{\lambda}{h_0} \ll Kn^*,
\label{eq:condKn}
\end{equation}
where $\lambda$ is the mean free path of the gas particles, and $Kn^*$ is a threshold value limiting our theory's regime of validity. It will be seen that ultimately the velocity profile plays a role only via the volumetric flux per unit length 
\begin{equation}
Q(x)  = \int\limits_{-h(x)}^{h(x)} v_x(x,z) dz.  \label{eq:Paper3_defQ}
\end{equation}
Previous experimental work in rectangular channels with $|\Delta P|/P_0 \approx 0.46$ shows that using the no-slip boundary conditions induces an error in the flow rate as little as $\approx 30\%$ for $Kn=0.1$ \cite{Colin2004}. We thus suggest $Kn^* = 0.1$. The mean free path may be computed using 
\begin{equation}
\lambda = \left(\frac{\pi}{8}\right)^{1/2} \frac{\eta}{u} \left( \frac{k_BT}{m}\right)^{1/2} \frac{1}{P},
\label{eq:formulaLambda}
\end{equation}
where $u$ is a numerical constant $\approx 0.5$ \cite{Jennings1988}. For air at $T \approx 300K$ and $P= 1$ atm, the above formula yields $\lambda \approx 10^{-2} \mu m$, which plugged into Eq \eqref{eq:condKn} yields $Kn \approx 10^{-2}$ when assuming $h_0 = 1 \mu m$. More generally, plugging Eq. \eqref{eq:formulaLambda} into Eq \eqref{eq:condKn} yields a condition on the minimum pressure for which the theory is still valid
\begin{equation}
P > \left(\frac{\pi}{8}\right)^{1/2} \frac{\eta}{u Kn^* h_0} \left( \frac{k_BT}{m}\right)^{1/2}, 
\end{equation}
which for air around $300 K$ ($\eta \approx 10^{-5} \text{Pa}\cdot\text{s}$ \cite{Nagashima1985}, $m \approx 5 \times 10^{-26} Kg$ \cite{Nagashima1985}) flowing in a channel of $h_0 = 1 \mu m$ yields $P \gtrsim 0.4 atm$.
As a summary, we now repeat all of the assumptions that define the regime of validity of our model:
\begin{equation*}
    \begin{cases}
      B_1 \bar{\rho}   \ll 1 & \text{ideal gas equation of state} \\
      \frac{h_0}{L}    \ll 1 & \text{lubrication approximation}\\
      \begin{rcases}
      Re \frac{h_0}{L} \ll 1 \\
      \frac{\zeta}{\eta} \left( \frac{h_0}{L} \right)^2 \ll 1 \\
      \end{rcases} &\text{Stokes equation}\\
      1 \ll Pe_x &\text{advection dominates transport in } x \\
      Pe_x \ll \left( \frac{L}{h_0} \right)^2  &\text{diffusion dominates transport in } z \\
      Da_z \ll 1 &\text{homogeneous density along } z\\
       Kn \ll Kn^* &\text{no-slip boundary conditions }
    \end{cases}
\end{equation*}
We now calculate  the flux of product at the reactor outlet.

\section{The yield of a catalytic gas-phase reactor}

\label{sec:yield}
We now have all the ingredients necessary to solve for the number densities and fluxes. Inserting Eq.~\eqref{eq:Paper3_v_x_MAIN} in Eqs.~\eqref{eq:Paper3_fluxA_lon_MAIN} and \eqref{eq:Paper3_fluxB_lon_MAIN}, we obtain the fluxes in the $x$ direction, which can then be plugged into Eq.~\eqref{eq:Paper3_constTotFlux}. This procedure yields
\begin{equation}
\partial_{x}\left[\rho(x)\int\limits_{-h(x)}^{h(x)}\frac{z^{2}-h^{2}(x)}{2\eta}\partial_{x}P(x)dz\right] = 0, \label{eq:Paper3_intermediateJ}
\end{equation}
which is solved by 
\begin{align}
\label{eq:Paper3_totRho} 
\rho(x) = \Bigg[&\rho(0)^2 -  [\rho(0)^2 - \rho(L)^2] \times \nonumber \\
& \times \left(  \int\limits_0^x h^{-3}(x) dx \right)\bigg/ \left( \int\limits_0^L h^{-3}(x) dx \right) \Bigg]^{1/2}
\end{align}
from which the pressure $P(x)$ is obtained, when plugged into the equation of state (Eq.~\eqref{eq:Paper3_idealGas}). The dependence of the number density in the longitudinal direction is entirely handled by the first term inside brackets of Eq.~\eqref{eq:Paper3_totRho}. 

We now compute the volumetric flux rate by combining Eq. \eqref{eq:Paper3_defQ} together with Eq.~\eqref{eq:Paper3_v_x_MAIN} is written as
\begin{equation}
Q(x)  = \partial_x P(x)\int\limits_{-h(x)}^{h(x)} \frac{z^{2}-h^{2}(x)}{2\eta} dz, 
\end{equation}
yielding
\begin{equation}
Q(x) = \frac{1}{3} \frac{P_0^2-P_L^2}{\eta k_BT} \left( \int\limits_0^L h^{-3}(x) dx \right)^{-1} \frac{1}{\rho(x)}. \label{eq:Paper3_Q}
\end{equation}
It can be seen from Eqs.~\eqref{eq:Paper3_defJ}, \eqref{eq:Paper3_fluxA_lon_MAIN} and \eqref{eq:Paper3_fluxB_lon_MAIN} that
\begin{equation}
J= \rho(x) Q(x), \label{eq:JandQ}
\end{equation}
which together with Eq.~\eqref{eq:Paper3_Q} yields 
\begin{equation}
    J = \frac{1}{3} \frac{P_0^2 -P^2_L}{\eta k_B T} \frac{h_0^3}{L}  \left[ \frac{1}{L} \int\limits_0^L \left(\frac{h(x)}{h_0}\right)^{-3}dx \right]^{-1}, 
\end{equation}
which can be rewritten as
\begin{equation}
    J = - \frac{2}{3}\frac{ P_0  h_0^3 }{\eta k_B T} \left[ \frac{1}{L} \int\limits_0^L \left(\frac{h(x)}{h_0}\right)^{-3}dx \ \right]^{-1}  \frac{\Delta P}{L} \left( 1 + \frac{1}{2}\frac{\Delta P}{P_0} \right),  \label{eq:Paper3_J} 
\end{equation}
after defining the pressure drop
\begin{equation}
    \Delta P = P_L - P_0.
\end{equation}
Henceforth, we will consider the pressure to be higher in the left ($x<0$) reservoir than in the right ($x>L$) reservoir, and thus $\Delta P <0$. This is done merely to facilitate the discussion by fixing the signs of the fluxes. Interchanging $P_0$ with $P_L$ ($\Delta P \rightarrow - \Delta P$ and $P_0 \rightarrow P_0 + \Delta P$) merely results in a sign switch for $J$ and $J_B(L)$. It is noteworthy that Eq.~\eqref{eq:Paper3_J} includes also a quadratic contribution in $\Delta P$. 

As chemical species $A$ and $B$ exhibit the same physical properties, the total fluid motion is insensitive to the composition of the mixture and hence both the total flow $J$ and total number density $\rho(x)$ show no dependence on the reaction rate $\alpha(x)$. Nonetheless, we are interested in maximizing the flow of product, and as such, we require the solution of Eqs.~\eqref{eq:Paper3_ADE_A} and \eqref{eq:Paper3_ADE_B}. Employing the same technique as for the total number density, we integrate Eq.~\eqref{eq:Paper3_ADE_A} along the $z$ direction to obtain
\begin{equation}
\int\limits_{- \infty}^{ \infty} \left[ \partial_x \bm{j}_{A,x}(x,z) + \partial_z \bm{j}_{A,z}(x,z) \right]dz = - \int\limits_{- \infty}^{ \infty} \xi(x,z) dz, \label{eq:Paper3_contSteady1_A}
\end{equation}
which reduces to 
\begin{equation}
 \partial_x \int\limits_{- \infty}^{ \infty}  \bm{j}_{A,x}(x,z) dz = -2\alpha(x) \rho_A(x), \label{eq:Paper3_contSteady2_A}
\end{equation}
as there is no flux of reactant at infinity, with $\xi(x,z)$ being given by Eq. \eqref{eq:Paper3_xi}. Indeed, as there is no flux anywhere outside the reactor (Eq.~\eqref{eq:Paper3_nofluxAoutside}), we have 
\begin{equation}
 \partial_x \int\limits_{-h(x)}^{h(x)}  \bm{j}_{A,x}(x,z) dz = -2\alpha(x) \rho_A(x), 
\end{equation}
By using Eqs. \eqref{eq:Paper3_fluxA_lon_MAIN} and \eqref{eq:Paper3_Aconst}, we write
\begin{equation}
\partial_{x}\left[\rho_A(x)\int\limits_{-h(x)}^{h(x)} v_x (x,z) dz\right] =  -2\alpha(x) \rho_A(x),
\end{equation}
and using the definition in Eq. \eqref{eq:Paper3_defQ}, we write
\begin{align}
\partial_x \left[ \rho_A(x) Q(x) \right] = -2\alpha(x) \rho_A(x),
\end{align}
which is solved by
\begin{equation}
\rho_A(x) =  \rho(x) \exp\left[-2 \frac{1}{L} \int\limits_0^x \frac{\alpha(x) L}{Q(x)} dx \right], 
\end{equation}
and using Eq.~\eqref{eq:Paper3_defTotRho} the density profile of products reads
\begin{equation}
\rho_B(x) = \rho(x) \left\{ 1 - \exp\left[-\frac{1}{L} \int\limits_0^L \frac{ 2\alpha(x) L}{Q(x)} dx \right] \right\}. 
\label{eq:rho_b}
\end{equation}
Similarly to Eq.~\eqref{eq:JandQ}, the flow of product per unit length is defined as
\begin{equation}
J_B(x)=\rho_B(x) Q(x), \label{eq:JBandQ}
\end{equation}
which at the outlet, using Eq.~\eqref{eq:rho_b}, reads
\begin{equation}
J_B(L) = J \left\{ 1 - \exp\left[-\frac{1}{L} \int\limits_0^L \frac{ 2\alpha(x) L}{Q(x)} dx \right] \right\}.  \label{eq:Paper3_JBoutlet}
\end{equation}
We now identify a local timescale associated with transport across the reactor $\tau_T(x)$
\begin{equation}
    \tau_T(x) = L \left( \frac{Q(x)}{2h(x)} \right)^{-1},
\end{equation}
and a local timescale associated to the reaction $\tau_R(x)$
\begin{equation}
    \tau_R(x) = \frac{h(x)}{\alpha_0(x)},
\end{equation}
which is the time associated with transport in the transverse direction with velocity $\alpha_0$. We identify the ratio of these timescales as the local Damk\"ohler number in the longitudinal direction 
\begin{equation}
    Da_x(x) = \frac{\tau_T(x)}{\tau_R(x)} = \frac{2\alpha(x) L}{Q(x)},  \label{eq:dammkohlerDef}
\end{equation}
such that
\begin{equation}
J_B(L) = J \left\{ 1 - \exp\left[- \frac{1}{L} \int\limits_0^L Da_x(x) dx \right] \right\}.  \label{eq:Paper3_J_B} 
\end{equation} 

Another variable of interest is the purity $\Pi$ at the outlet, defined as the ratio of product density to total density:
\begin{equation}
    \Pi = \frac{\rho_B(L)}{\rho(L)}  \label{eq:purity_def} 
\end{equation}
By using Eqs.~\eqref{eq:JandQ} and \eqref{eq:JBandQ}, we write
\begin{equation}
    \Pi = \frac{J_B(L)}{J}, \label{eq:purity_def2} 
\end{equation}
and using Eq.~\eqref{eq:Paper3_JBoutlet},
\begin{equation}
    \Pi = 1 - \exp\left[- \frac{1}{L} \int\limits_0^L Da_x(x) dx \right],  \label{eq:purity} 
\end{equation}
which depends only on the spatially-averaged Damk\"ohler number (in the longitudinal direction). In the following sections, we analyze the equations we have just derived in the nearly incompressible limit.

\section{The role of compressibility}

\label{sec:incomp}
In this section, we shed light on the role of compressibility on the output of a catalytic microreactor. Indeed, it is known that compressible fluids behave close to being incompressible when weakly driven, i.e., when the density varies only a little around an average value. We now show that our equations describe an incompressible fluid when this limit is taken. As such, we are able to describe both liquid-phase and gas-phase reactors and are able to pinpoint the role the fluid phase plays in the reactor output. As discussed, we study Eqs.~\eqref{eq:Paper3_J} and \eqref{eq:Paper3_J_B} in the limit of weak density variations. Let the density $\rho(x)$ be written as
\begin{equation}
    \rho(x) = \rho(0) + \delta \rho(x),
\end{equation}
where $\delta \rho(x)$ is the variation of density with respect to the inlet value. We now focus on the limit
\begin{equation}
    \frac{ \delta \rho(x) }{ \rho(0) } \ll 1 \label{eq:incompLimit}
\end{equation}
and perform expansions of the fluxes in this small parameter, keeping only terms up to linear order. The total flow $J$ in Eq.~\eqref{eq:Paper3_J} yields
\begin{equation}
    J \approx - \frac{2}{3}\frac{ P_0  h_0^3 }{\eta k_B T} \left[ \frac{1}{L} \int\limits_0^L \left(\frac{h(x)}{h_0}\right)^{-3}dx \ \right]^{-1}  \frac{\Delta P}{L} 
\end{equation}
and by using the equation of state of Eq.~\eqref{eq:Paper3_idealGas}, we rewrite the above equation as
\begin{equation}
    J \approx - \frac{2}{3}\frac{ \rho(0)  h_0^3 }{\eta} \left[ \frac{1}{L} \int\limits_0^L \left(\frac{h(x)}{h_0}\right)^{-3}dx \ \right]^{-1}  \frac{\Delta P}{L}. \label{eq:J_liquid}
\end{equation}
We also obtain the volumetric flux $Q$ to linear order from Eq.~\eqref{eq:Paper3_Q}:
\begin{equation}
    Q \approx -\frac{2}{3} \frac{\Delta P h_0^3}{L \eta} \left[ \frac{1}{L} \int\limits_0^L \left(\frac{h(x)}{h_0}\right)^{-3}dx \ \right]^{-1} , 
\end{equation}
which does not vary in space. As such, Eqs.~\eqref{eq:Paper3_J} and \eqref{eq:Paper3_J_B} to linear order in $\delta \rho(x)/ \rho(0)$ describe the flow of an incompressible fluid, such as a liquid. Finally, we may obtain the flow of product at the outlet from Eq.~\eqref{eq:Paper3_JBoutlet} as
\begin{equation}
    J_B(L) = J \left\{ 1 - \exp\left[-\frac{2 \alpha_0 L}{Q}\right] \right\}, \label{eq:JB_liquid}
\end{equation}
where only $\alpha_0$ enters the final result due to $Q$ being homogeneous in space. As such, the output of a liquid-phase reactor is insensitive to spatial variations in the catalytic concentration. From comparing Eq.~\eqref{eq:JB_liquid} to Eq.~\eqref{eq:Paper3_J_B}, we identify the average Damk\"ohler number for liquids as
\begin{equation}
    \langle Da \rangle_x = \frac{2 \alpha_0  L}{Q}. \label{eq:Dax_liq}
\end{equation}
The purity (defined in Eq. \eqref{eq:purity_def}) may be obtained from Eqs.~\eqref{eq:purity_def2} and  \eqref{eq:JB_liquid} as 
\begin{equation}
    \Pi =  1 - \exp\left[-\frac{2 \alpha_0 L}{Q}\right]. 
\end{equation}
We now relax the incompressibility and focus on the role of the shape of the reactor, as well as the distribution of catalytic material inside it. To do so, we apply our theory to the case of a sinusoidal reactor shape and catalytic coating density. Such functional forms are mathematically simple, but still capture the general effect of introducing corrugation to the shape and inhomogeneities to the catalytic coating.

\begin{figure*}[t]
	\centering
   \includegraphics[width=1.0\textwidth]{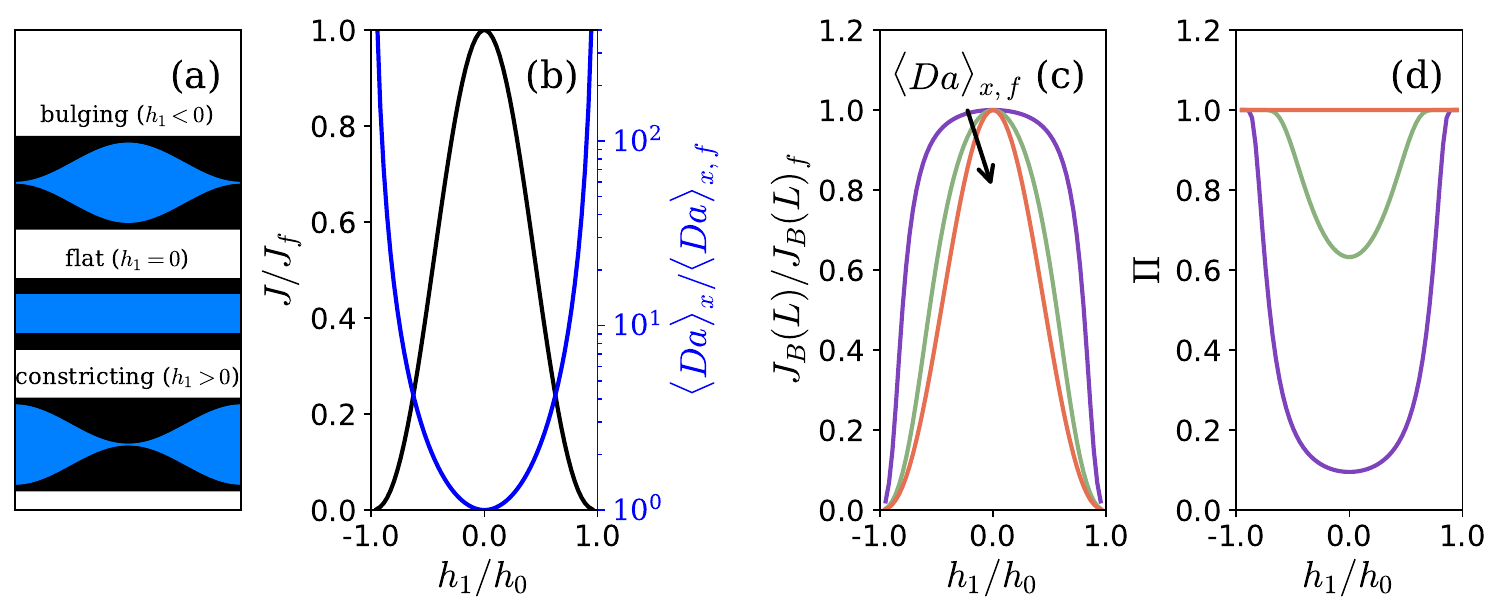}
   \caption{Panel (a): sketches of three different reactor shapes. Black regions indicate the reactor wall and blue regions indicate the reactor chamber filled with gas. Panel (b): black line: total flux $J$ normalized by value for a flat ($h_1=0$) reactor $J_f$ (values reported on the left axis). \textcolor[HTML]{0000FF}{Blue} line:  Damk\"ohler number $\langle Da \rangle_x$ normalized by value for a flat reactor $\langle Da \rangle_{x,f}$ (values reported on the right axis). Panel (c): flow of product at reactor outlet $J_B(L)$ normalized by value for a flat reactor $J_B(L)_f$. Panel (d): purity at reactor outlet $\Pi$. For panels (c) and (d): $\langle Da \rangle_{x,f} \in \{10 \text{ (\textcolor[HTML]{E76F51}{orange})} ,1\text{ (\textcolor[HTML]{8AB17D}{green})}, 10^{-1} \text{ (\textcolor[HTML]{7E43BC}{violet})}\}$. 
   }
   \label{fig:liquid}
\end{figure*}

\section{Sinusoidal reactor shape and catalytic coating density}

In this section, we wish to investigate how the flow rate of product may be optimized by a suitable choice of reactor shape, or of catalytic coating. To this end, we now apply our theory to the concrete case of a reactor with a sinusoidal shape, 
\begin{equation}
    h(x) = h_0 + h_1 \cos\left( k x\right) , \label{eq:Paper3_h_cos} 
\end{equation}
where $h_1$ is a constant determining the height of the corrugation and 
\begin{equation}
k = \frac{2 \pi}{L}.
\end{equation} 
Note that the reactor shape is fore-aft symmetric. For values of $h_1 < 0$, the reactor is narrow at the inlet/outlet and bulges in the middle. Conversely, for values $h_1 > 0$, the channel is wide at the inlet/outlet and narrows in the middle. The value $h_1=0$ corresponds to a flat reactor. See Fig.~\ref{fig:liquid}(a) for a sketch of the aforementioned reactor geometries.  We further take a sinusoidal reaction rate
\begin{equation}
    \alpha(x) = \alpha_0 + \alpha_1 \sin\left( k x\right), \label{eq:Paper3_alpha_cos} 
\end{equation}
with $\alpha_0$ being the average reaction rate, and $\alpha_1$ determining the deviation from that average. Note that the reaction rate is in general not fore-aft symmetric, with the sign of $\alpha_1$ determining if more catalytic material is found near the inlet ($\alpha_1 >0$) or near the outlet ($\alpha_1 <0$). This choice of $h(x)$ and $\alpha(x)$ is simple, yet captures the general effect of corrugation and inhomogeneities in the catalytic coating.

\subsection{Incompressible fluid reactors}

The simplest case is that of incompressible flows (such as a liquid-phase reactor), as the fluid density is homogeneous. The effect of corrugations on the total flux $J$ and in the Damk\"ohler number $\langle Da \rangle_x$ can be seen from Eqs.~\eqref{eq:Paper3_J} and \eqref{eq:Dax_liq}. Indeed, one may write 
\begin{equation}
    \frac{J}{J_f} = \left[ \frac{1}{L} \int\limits_0^L \left(\frac{h(x)}{h_0}\right)^{-3}dx \ \right]^{-1} \label{eq:Paper3_J_Jf}
\end{equation}
and
\begin{equation}
    \frac{\langle Da\rangle_x}{\langle Da\rangle_{x,f}}  =  \frac{1}{L} \int\limits_0^L \left(\frac{h(x)}{h_0}\right)^{-3}dx , \label{eq:Paper3_Da_Daf_incomp}
\end{equation}
where $J_f$ and $\langle Da \rangle_{x,f}$ are the respective values of $J$ and $\langle Da \rangle_x$ for flat channels ($h_1 = 0$).  Note that Eq.~\eqref{eq:Paper3_J_Jf} remains valid even for non-sinusoidal reactor shapes. Using Eq.~\eqref{eq:Paper3_h_cos}, the integral on the right-hand-side of Eq.~\eqref{eq:Paper3_J_Jf} can be computed numerically, leading to the results of Fig.~\ref{fig:liquid}(b). In this Figure, we see that the net effect of the corrugation ($h_1 \neq 0$) is to decrease the total flux, with a corresponding increase in the Damk\"ohler number.  This decrease in $J$ is due to the presence of a bottleneck that becomes narrower the larger the value of $|h_1|/h_0$. Indeed, for $|h_1|/h_0 = 1$ the bottleneck is shut, and no gas can flow through ($J=0$). Furthermore, $J/J_f$ is insensitive to the sign of $|h_1|/h_0$. As a result, bulging reactors ($h_1 <0$) exhibit the same total flow as constricting reactors ($h_1 >0$). An analysis of Eq.~\eqref{eq:JB_liquid} shows that the flow rate of product $J_B(L)$ normalized by its value for flat channels depends only on the numbers $h_1/h_0$ and $\langle Da \rangle_{x,f}$. We plot this data in Fig.~\ref{fig:liquid}(c), where it can clearly be seen that the shape that maximizes the flow rate of product is the flat reactor for all values of $\langle Da \rangle_{x,f}$. The latter number controls only the width of the peak around $h_1=0$. Conversely, Fig.~\ref{fig:liquid}(d) shows that, while flat channels exhibit the largest product fluxes, they also lead to the lowest purity. The large flow velocities that lead to high $J_B(L)$ also decrease the time that reactant particles stay inside the reactor and, as such, the time they have to react with the catalytic wall. We now relax the condition of incompressibility and study a general reactor with homogeneous catalytic coating.

\subsection{Compressible fluid reactors - homogeneous catalytic coating ($\alpha_1 = 0$)}

\begin{figure}[t]
 \begin{minipage}{\linewidth}
 \begin{center}
 \includegraphics[width=1.0 \linewidth]{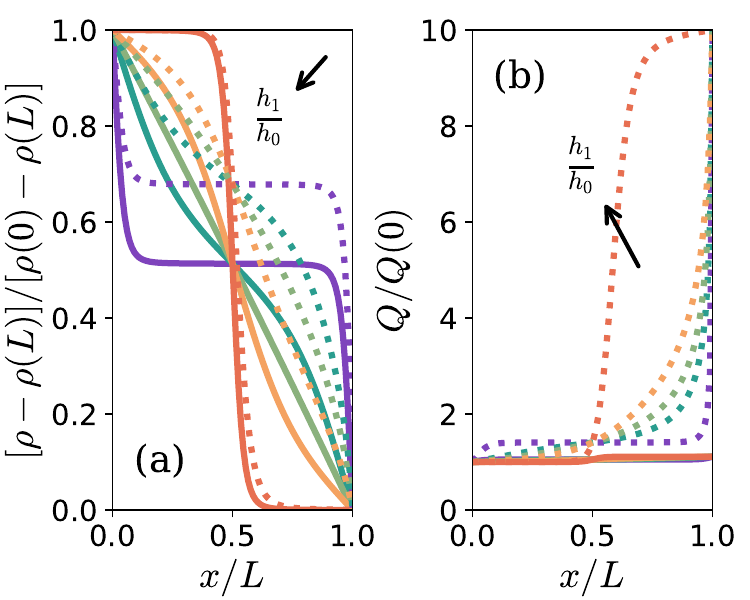} 
 \end{center}
  \caption{ Panel (a): difference between gas density $\rho(x)$  and its value at outlet $\rho(L)$ (normalized by difference between inlet and outlet density, $\rho(0)$ and $\rho(L)$, respectively) as a function of position for sinusoidal reactors of different corrugation parameters $h_1/h_0$. Panel (b): Volumetric flow rate $Q(x)$ normalized by inlet value $Q(0)$. For all panels $|\Delta P|/P_0 \in \{0.1 (\full),  0.9 (\dotted) \}$, and $h_1/h_0 \in \{-0.9,-0.2,0,0.2,0.9\}$. }\label{fig:Gas_HomoReaction} \end{minipage}
\end{figure}
\begin{figure}[t]
 \begin{minipage}{\linewidth}
 \begin{center}
 \includegraphics[width=1.0 \linewidth]{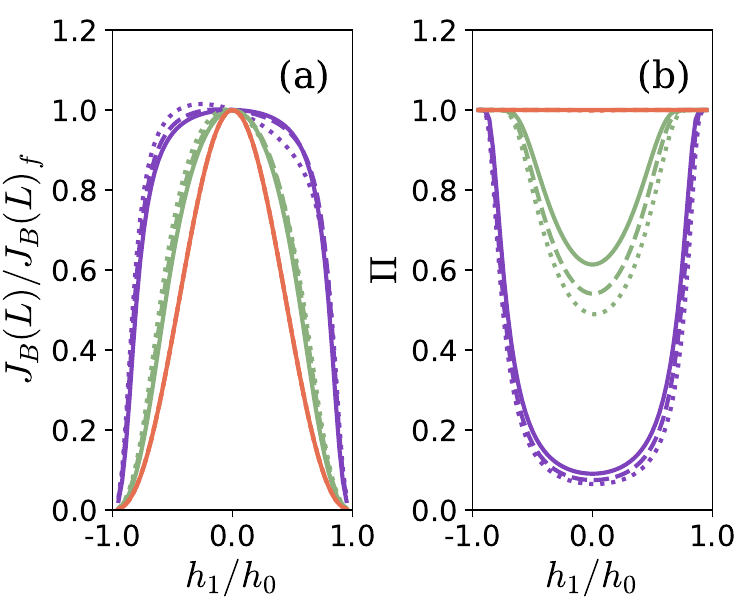} 
 \end{center}
  \caption{ Panel (a): Flow of product at reactor outlet $J_B(L)$ normalized by value for a flat reactor $J_B(L)_f$. Panel (b): Purity at outlet $\Pi$. For both panels: $|\Delta P|/P_0 \in \{0.1 (\full), 0.5(\dashed), 0.9 (\dotted) \}$, $\alpha_1/\alpha_0 = 0$, and $ 2\alpha_0 L/ Q_f(0) \in \{10 \text{ (\textcolor[HTML]{E76F51}{orange})} ,1\text{ (\textcolor[HTML]{8AB17D}{green})}, 10^{-1} \text{ (\textcolor[HTML]{7E43BC}{violet})}\}$. The function $Q_f(x)$ is defined as $Q(x)$ for $h_1 = 0$.   }
    	\label{fig:Gas_HomoReaction_Fluxes} \end{minipage}
\end{figure}

In this section, we study compressible flow reactors. For clarity of discussion, we start with the relatively simple case of homogeneous catalytic coating. The first major difference between the incompressible and compressible regime is that Eq.~\eqref{eq:Paper3_Da_Daf_incomp} is not valid for the latter, as $Q(x)$ (and $\rho(x)$) is now spatially varying. According to Eq.~\eqref{eq:dammkohlerDef}, we have
\begin{equation}
     \langle Da \rangle_x = 2 \alpha_0 L \langle Q^{-1}(x) \rangle. 
\end{equation}
To analyze the behavior of $\langle Q^{-1}(x) \rangle$, it is useful to first inspect the spatial variation of both the volumetric flow rate and the density.  We may now combine Eqs.~\eqref{eq:Paper3_totRho} and \eqref{eq:Paper3_h_cos} to numerically obtain the density profile inside the reactor, which we plot in Fig.~\ref{fig:Gas_HomoReaction}(a). As expected, the presence of corrugations ($h_1 \neq 0$) leads to density/pressure profiles that decrease in a sharper fashion near the bottleneck, which is located at $z = L/2$ for constricting channels ($h_1 > 0$), and at $z \in \{0,L\}$ for bulging channels ($h_1 <0$). This behavior is seen for both low relative pressure drops ($|\Delta P|/P_0 = 0.1$) and high relative pressure drops ($|\Delta P|/P_0 = 0.9$ ). Comparing these two plots for the case of flat channels ($h_1 = 0$), one clearly sees the effect of compressibility. For low relative pressure drops, the pressure profile is nearly a straight line, as expected from incompressible flows in constant section channels. However, the pressure profile for constant section channels becomes curved at high relative pressure drops. The reason for this non-linearity is that the density profile controls not only the amount of gas inside the reactor but also the pressure profile and, thus the flow velocity. Therefore, the condition for steady state (i.e. homogeneous total flux $J$) is non-linear in $\partial_x \rho(x)$ (see Eq.~\eqref{eq:Paper3_intermediateJ}). In particular, to keep the total flow $J$ constant in space, regions with smaller gas density require larger pressure gradients (in magnitude). One clearly sees that this non-linearity in $\partial_x \rho(x)$ leads to a spatially-averaged density $\langle \rho(x) \rangle$ that changes with $h_1$ at high relative pressure drops. In particular, negative values of $h_1$ showcase higher values of $\langle \rho(x) \rangle$. Finally, note that all pressure profiles intersect at $x/L=0.5$. This is a result of the fore-aft symmetry of the reactor shape $h(x)$, as defined in Eq.~\eqref{eq:Paper3_h_cos}.

Having obtained the density profile, we may obtain the volumetric flow rate $Q(x)$ from Eq.~\eqref{eq:Paper3_Q}, which we plot in Fig.~\ref{fig:Gas_HomoReaction}(b). Indeed, the latter and the former are inversely proportional to one another, as per Eq.~\eqref{eq:JandQ}, and so all previous reasoning about the density profile applies to the volumetric flux. As with the gas density, the spatially-averaged volumetric flow rate $\langle Q(x) \rangle$ is also dependent on $h_1$ and is sensitive to its sign. Therefore, reactant particles spend more time inside bulging reactors ($h_1 <0$) than inside constricting reactors ($h_1 <0$). Such an effect implies a breaking of symmetry around $h_1=0$ for $J_B(L)$ and $\Pi$, as we see in Figs.~\ref{fig:Gas_HomoReaction_Fluxes}. While this effect is small (and for these parameters, only visible for $|\Delta P|/P_0 = 0.9$ and low purity reactors), it shows that the flow rate of corrugated reactors can outperform that of flat reactors. Next, we introduce an inhomogeneity in the catalytic coating in an attempt to increase this gain.

\subsection{Inhomogeneous catalytic coating ($\alpha_1 \neq 0$) }

\label{sec:inhomo}

\begin{figure*}[t]
	\centering
   \includegraphics[width=1.0\textwidth]{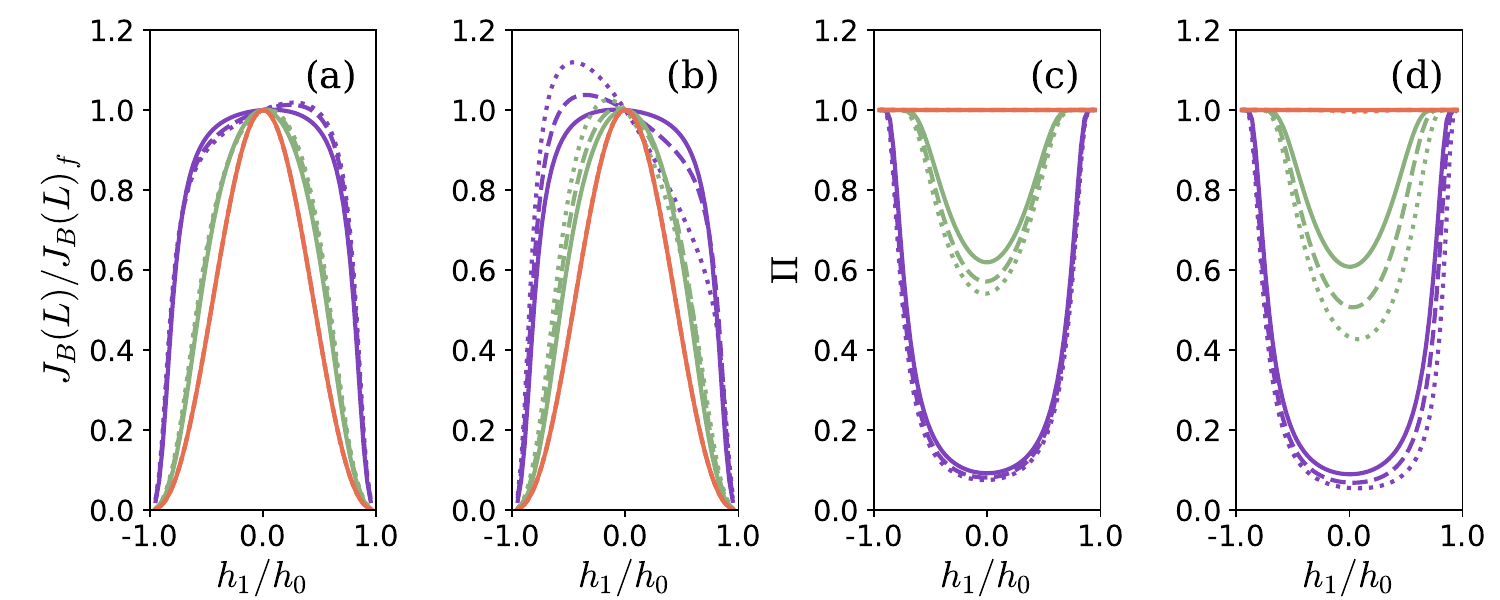}
   \caption{ Flow of product at reactor outlet $J_B(L)$ normalized by value for a flat reactor $J_B(L)_f$. For panel (a): $\alpha_1/\alpha_0 = 0.9$, and for panel (b): $\alpha_1/\alpha_0 = -0.9$. For all panels: $|\Delta P|/P_0 \in \{0.1 (\full), 0.5(\dashed), 0.9 (\dotted) \}$, and $ 2\alpha_0 L/ Q_f(0)  \in \{10 \text{ (\textcolor[HTML]{E76F51}{orange})} ,1\text{ (\textcolor[HTML]{8AB17D}{green})}, 10^{-1} \text{ (\textcolor[HTML]{7E43BC}{violet})}\}$. The function $Q_f(x)$ is defined as $Q(x)$ for $h_1 = 0$. \label{fig:Paper3_h1_inhomo}}
\end{figure*}

\begin{figure}[t]
	\centering
   \includegraphics[width=0.5\textwidth]{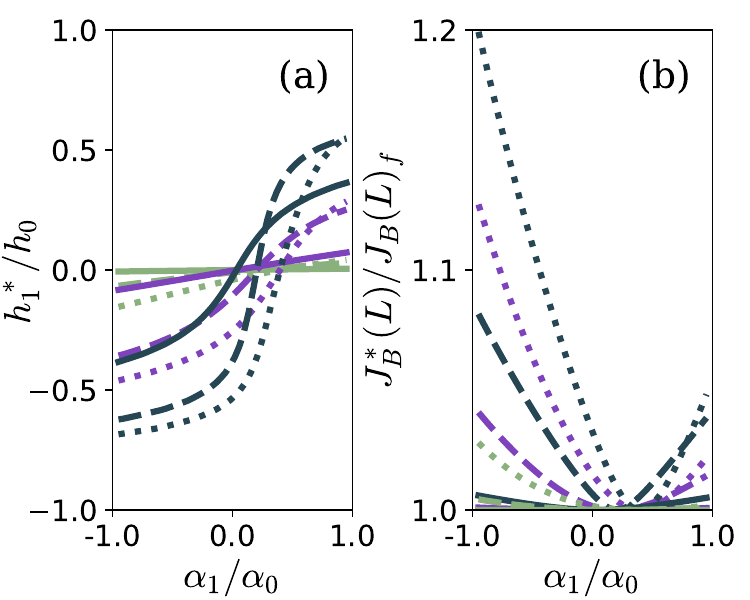}
   \caption{Panel (a): Optimum value of the corrugation parameter $h_1^*/h_0$. Panel (b): Optimum flow rate of product at outlet $J_B^*(L)$ normalized by value $J_B(L)_f$ for flat reactor ($h_1 = 0$). Parameters are $-\Delta P/P_0 \in \{ 0.1 (\full), 0.5 (\dashed), 0.9 (\dotted)\}$, and $2\alpha_0 L/ Q_f(0) \in \{1 \text{ (\textcolor[HTML]{8AB17D}{green})} ,10^{-1} \text{ (\textcolor[HTML]{7E43BC}{violet})}, 10^{-2} \text{ (\textcolor[HTML]{264653}{gray})}\}$. The function $Q_f(x)$ is defined as $Q(x)$ for $h_1 = 0$.
   \label{fig:max_h1_JB_alpha1}}
\end{figure}

In this section, we study compressible flow reactors with inhomogeneous catalytic coatings ($\alpha_1 \neq 0$). According to Eq.~\eqref{eq:dammkohlerDef}, we have
\begin{equation}
     \langle Da \rangle_x = 2 \alpha_0 L \left( \langle Q^{-1}(x) \rangle + \frac{\alpha_1}{\alpha_0}  \langle \sin(x) Q^{-1}(x) \rangle \right) . \label{eq:Daz_compressible_inhomo}
\end{equation}
Because $Q(x)$ is a monotonically growing function (see Fig.~\ref{fig:Gas_HomoReaction}(b)), the last term in Eq.~\eqref{eq:Daz_compressible_inhomo} is largest when $\alpha_1 = 1$, keeping all other parameters unchanged. Thus, a reactor operating at a large relative pressure drop showcases the highest yields (both flow of product and purity) when the catalytic material is deposited closest to the inlet. Near the inlet, the flow velocities take their smallest values, and so reactant particles spend the most time there. However, the case may arise where the catalyst distribution is more difficult to tune experimentally than the reactor corrugation. In that case, one observes that there is an optimum corrugation parameter as in the homogeneously coated case, but which now depends on the catalytic distribution (Figs.~\ref{fig:Paper3_h1_inhomo}(a) and (b)). Indeed, Fig.~\ref{fig:Paper3_h1_inhomo}(b) shows that an inhomogeneous distribution of catalytic material may boost the gain one obtains by introducing corrugations. Note that this gain is largest for the least optimal catalyst distributions ($\alpha_1 <0$), for high relative pressure drops, and for low values of $\Pi$  (Fig.~\ref{fig:Paper3_h1_inhomo}(c) and (d)). In Fig.~\ref{fig:max_h1_JB_alpha1}(a), we sweep over all values of $\alpha_1/\alpha_0$ and show how the optimum corrugation parameter depends on the catalyst distribution. In general, bulging reactors ($h_1 >0$) are best if the catalyst is most concentrated near the outlet ($\alpha_1 <0$). Conversely, constricting reactors ($h_1 >0$) are best if the catalyst is most concentrated near the inlet ($\alpha_1>0$). Finally, Fig.~\ref{fig:max_h1_JB_alpha1}(b) shows that the largest gains (with respect to flat reactors) are obtained when the catalyst distribution is furthest away from being homogeneous ($\alpha_1 = - 1$).

\section{Experimental results using catalytic membranes}
\begin{figure}[t]
	\centering
   \includegraphics[width=0.5\textwidth]{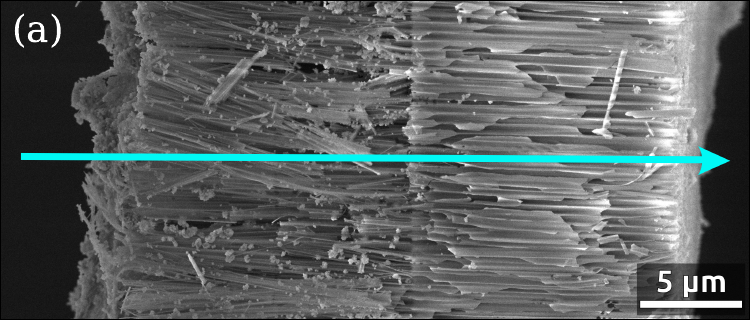}
   \includegraphics[width=0.5\textwidth]{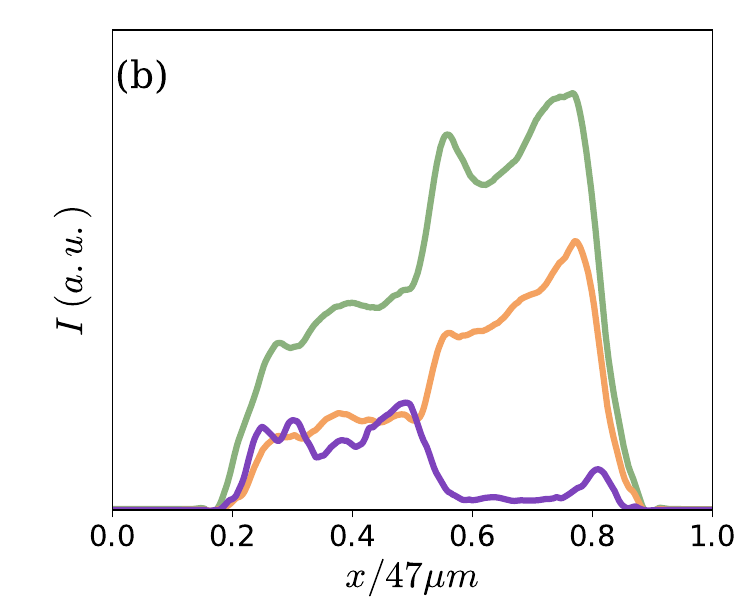}
   \caption{Characterization of a membrane coated with catalytic nanoparticles. Panel (a): Scanning electron micrograph of the membrane in cross-section view. Cyan arrow describes the paths scanned for the data in panel (b). Catalytic nanoparticles visible as lighter spheres peppering the inside of the membrane. Panel (b): Intensity $I$ (in arbitrary units) given by cross-section energy-dispersive X-ray spectroscopy analysis measured along the pores (see cyan arrow in panel (a)). Higher values of $I$ indicate larger concentrations of a given element: aluminium (\textcolor[HTML]{8AB17D}{green}), oxygen (\textcolor[HTML]{E76F51}{orange}), and gallium (\textcolor[HTML]{7E43BC}{violet}). The latter is a marker for the nanoparticles and acts as the catalytic material. 
   \label{fig:microscope}}
\end{figure}

\begin{figure}[t]
	\centering
   \includegraphics[width=0.5\textwidth]{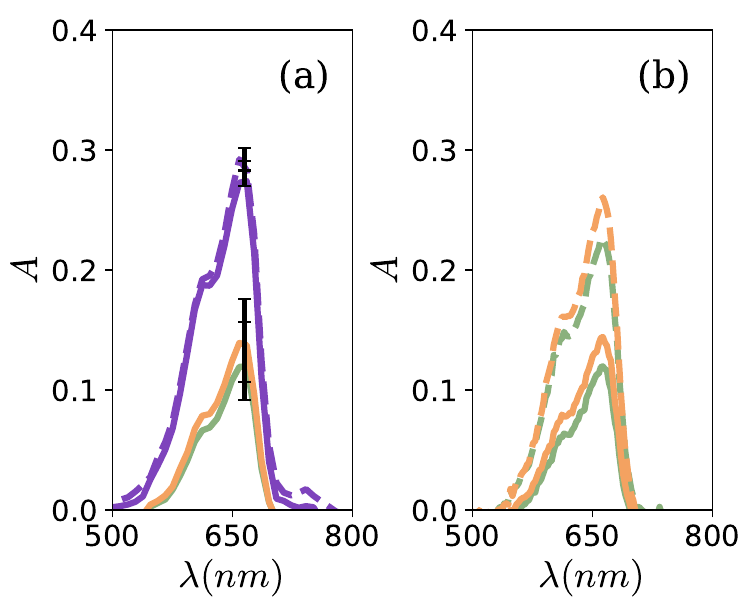}
   \caption{Data for absorbance $A$ as function of wavelength $\lambda$. Higher values of $A$ indicate larger concentrations of reactant. Panel (a): Violet solid line (\textcolor[HTML]{7E43BC}{\full}): measurement at a position upstream of the membrane. Violet dashed line (\textcolor[HTML]{7E43BC}{\dashed}): measurement at a position downstream of the membrane (no catalyst). Orange line (\textcolor[HTML]{E76F51}{\full}): measurement at a position downstream of the membrane (catalyst loaded preferentially near inlet). Green line (\textcolor[HTML]{8AB17D}{\full}): measurement at a position downstream of the membrane (catalyst loaded preferentially near the outlet).  Panel (b):  Dashed lines (\dashed): measurement at a position upstream of the membrane. Solid lines (\full): measurement at space downstream of the membrane. \textcolor[HTML]{8AB17D}{Green} lines: measurement at a time before flipping the membrane. \textcolor[HTML]{E76F51}{Orange} lines: measurement at a time after flipping the membrane)
   \label{fig:exps}}
\end{figure}

We performed experiments with catalytic membranes with heavily fore-aft asymmetric distribution of catalytic material (see Appendix \ref{sec:experiments} for experimental details). These experiments aimed at verifying the prediction stated in Sec.~\ref{sec:incomp} that incompressible flow reactors should exhibit the same flow rate of product no matter how catalyst material is distributed on the reactor walls, provided that the total catalyst mass is preserved (same $\alpha_0$). To verify this prediction, we performed a catalytic study using Anodic Aluminum Oxide membranes loaded with SCALMS nanoparticles which act as the catalyst. As seen in Fig.~\ref{fig:microscope}, the loading is heavily fore-aft asymmetric. We flowed a liquid solution of methylene blue and ascorbic acid (the reactants) through the membrane, which led to the former reducing to leuco-methylene blue (the product). By measuring the light absorbance of the fluid mixture at 665~nm, we were able to monitor the density of reactant in different positions of our experimental setup.  In Fig.~\ref{fig:exps}(a), we show that the membrane is indeed catalytically active as there is a decrease in absorbance when comparing samples up- and downstream of the membrane. Furthermore, this activity is due to the catalyst and not to the support itself, as unloaded membranes do not affect the absorbance (violet lines in Fig.~\ref{fig:exps}(a)). When measuring a sample downstream of membranes loaded preferentially near the inlet or outlet, no difference is seen within the error margin (orange and green lines in Fig.~\ref{fig:exps}(a)). To strengthen this conclusion, we performed an additional experiment with the same catalytic membrane, which, after some time, was flipped around so that the inlet and outlet switch positions. Again, no significant difference is seen in the absorbances.

\section{Conclusions}

In this paper, we analyzed the performance of an isothermal slit-shaped gas-phase reactor whose walls are coated with a catalytic material and through which flow is driven via an externally-imposed pressure drop. We derived a Fick-Jacobs theory, where the quasi-3D Navier-Stokes and advection-diffusion equations that govern the transport of the gaseous reactant and product were systematically reduced to an effective one-dimensional set of equations. These equations are applicable to corrugated reactors whose length far exceeds their average height and which are symmetric with respect to their centerline. Closed-form solutions for the pressure profile inside the reactor, the composition of the gas, and the flow of product and reactant are given for any corrugation that varies weakly.  Furthermore, the density of catalytic material is also taken as spatially-varying, and thus, so is the rate of the chemical reaction that turns reactant into product. 

To investigate the effect of corrugation and of inhomogeneous reaction rate, we studied a model reactor where both the reactor shape and reaction rate profile are sinusoidal. For small pressure drops relative to the inlet pressure, flat reactors exhibit the highest flow rate of product when compared with corrugated reactors of the same volume. In this incompressible limit, the reactor yield is insensitive to the distribution of catalytic material, which we verify experimentally. For compressible flows, we find an optimal non-flat reactor shape. This optimum shape is controlled by the inhomogeneity of the catalytic coating, as is the gain in product flow rate compared to a flat reactor. We find that reactors, where catalytic material is mostly located near the reactor inlet, can be optimized by a shape that is narrowest in the center. Conversely, a reactor where catalytic material is mostly found near the outlet can be optimized by a shape that is widest at the center.

Thus, one may optimize the output of a reactor via its shape or via an appropriate distribution of catalytic material rather than by the addition of further costly catalyst. We listed three reactions for which the assumptions of our model hold: decomposition of methane and carbon monoxide and production of ammonia. All three are of environmental or economic relevance \cite{Balcombe2017, Raub2000,Ghavam2021}. Finally, our results for the highly compressible regime may also be relevant for the design of novel low-pressure gas chromatography systems. These devices frequently operate with near-vacuum conditions at the outlet \cite{Zoccali2019,Jaap2000}, thus exhibiting values of relative pressure drop close to its maximum value.

\begin{acknowledgements}

We acknowledge funding by the Deutsche Forschungsgemeinschaft (DFG, German Research Foundation)—Project-ID 416229255—SFB 1411 and Project-ID 431791331—SFB 1452. Furthermore, we acknowledge the Helmholtz Association of German Research Centers (HGF) and the Federal Ministry of Education and Research (BMBF), Germany for supporting the Innovation Pool project ``Solar H2: Highly Pure and Compressed''. 

\end{acknowledgements}

\section*{Conflicts of Interest}
There are no conflicts of interest to declare.

\section*{Data Availability Statement}
The data that support the findings of this study is openly available in Zenodo at  \url{ http://doi.org/10.5281/zenodo.10161303 } 

\section*{Author Contributions}
GCA: conceptualization, investigation, formal analysis, and writing - original draft. MJ-S: conceptualization, investigation, formal analysis, writing - original draft. PM: conceptualization, formal analysis, supervision, and writing - review \& editing. JB: conceptualization, funding acquisition, supervision, writing - review \& editing. JH: conceptualization, funding acquisition, supervision, writing - review \& editing.

\appendix

\section{Systematic reduction to effective 1D equation}
\label{sec:1D}
 As stated in the main text, we now show how to solve Eq.~\eqref{eq:Paper3_constTotFlux} by a systematic reduction of the quasi-3D problem to an effective one-dimensional problem. We assume a long reactor such that the P\'eclet number $Pe_x$ along the $x$ direction is large,
\begin{equation}
    Pe_x = \frac{\overline{v}_x L}{D} \gg 1, \label{eq:Paper3_Pe_x}
\end{equation}
where $\overline{v}_x$ is the typical velocity in the $x$ direction. In such a reactor, advection is the dominant transport method in the longitudinal direction and
\begin{align}
\rho_A(x,z) v_x(x,z) &\gg \left( j_{D}^{(A)} \right)_x(x,z) \ ,\label{eq:Paper3_diffXsmall} \\
\rho_B(x,z) v_x(x,z) &\gg \left( j_{D}^{(B)} \right)_x(x,z) \ ,
\end{align}
where $\left( j_{D}^{(A)} \right)_x(x,z)$ and $\left( j_{D}^{(B)} \right)_x(x,z)$ are the $x$ component of the diffusive fluxes of $A$ and $B$ species, respectively. Thus, the fluxes in the $x$ direction are obtained from Eqs.~\eqref{eq:Paper3_flowA} and \eqref{eq:Paper3_flowA} as
\begin{align}
(\bm{j}_A )_x(x,z) &\approx \rho_A(x,z) \bm{v}_x(x,z), \label{eq:Paper3_fluxA_lon} \\
(\bm{j}_B )_x(x,z) &\approx \rho_B(x,z) \bm{v}_x(x,z), \label{eq:Paper3_fluxB_lon} 
\end{align}
where $(\bm{j}_A )_x(x,z)$ and $(\bm{j}_B )_x(x,z)$ are the $x$ components of fluxes of $A$ and $B$ particles, respectively. Plugging Eqs.~\eqref{eq:Paper3_fluxA_lon} and \eqref{eq:Paper3_fluxB_lon} in Eq.~\eqref{eq:Paper3_constTotFlux} yields
\begin{equation}
\partial_{x}\left[\int_{-h(x)}^{h(x)} \rho(x,z)\bm{v}_x(x,z) dz\right] = 0.
\end{equation}
We obtain $\bm{v}_x(x,z)$ from solving the Navier-Stokes equation (Eq.~\eqref{eq:Paper3_NS}) for the steady state
\begin{align}
    &0 = - \nabla P(x,z) + \eta \Delta \bm{v}(x,z) + \left (\zeta + \frac{1}{3}\eta \right) \nabla [\nabla \cdot \bm{v}(x,z)] - \nonumber \\
    &- \rho_f(x,z)  [\bm{v}(x,z) \cdot \nabla]  \bm{v}(x,z), \label{eq:Paper3_Stokes} 
\end{align}
the $x$ and $z$ component of which are
\begin{align}
0 = - &\partial_x P(x,z) + \eta \left[\partial_x^2 v_x (x,z) + \partial_z^2 v_x(x,z)\right] + \nonumber \\
      &\left(\zeta + \frac{1}{3}\eta \right)\partial_x \left[\partial_x v_x (x,z) + \partial_z v_z (x,z)\right] - \nonumber \\
      &-\rho_f(x,z) \left[ v_x(x,z) \partial_x v_x(x,z) + v_z(x,z) \partial_z v_x(x,z) \right],\label{eq:Paper3_xComponentStokes} \\
0 = - &\partial_z P(x,z) + \eta \left[\partial_x^2 v_z (x,z) + \partial_z^2 v_z(x,z)\right] + \nonumber \\
&\left(\zeta + \frac{1}{3}\eta \right)\partial_z \left[\partial_x v_x (x,z) + \partial_z v_z (x,z)\right] - \nonumber \\
& - \rho_f(x,z) \left[ v_x(x,z) \partial_x v_z(x,z) + v_z(x,z) \partial_z v_z(x,z) \right] , \label{eq:Paper3_yComponentStokes}
\end{align}
respectively. We now restrict ourselves to the case of a thin reactor,
\begin{equation}
\frac{h_0}{L} \ll 1, \label{eq:Paper3_thinreactor}
\end{equation}
which greatly simplifies the solution of Eqs.~\eqref{eq:Paper3_xComponentStokes} and \eqref{eq:Paper3_yComponentStokes}, as we now show. Let us normalize the $x$ and $z$ coordinates by the reactor length and average height, respectively. As such, we introduce the dimensionless variables $\widehat{x}$ and $\widehat{z}$ such that
\begin{align}
\widehat{x} &= \frac{x}{L}, \\
\widehat{z} &= \frac{z}{h_0}.
\end{align}
Furthermore, we define the typical velocity scale $\overline{v}_z$ in the $z$ direction. As such, we introduce the dimensionless velocities
\begin{align}
\widehat{v}_x(x,z) &= \frac{v_x(x,z)}{\overline{v}_x}, \\
\widehat{v}_z(x,z) &= \frac{v_z(x,z)}{\overline{v}_z},
\end{align}
which are of order unity. Similarly, we introduce the typical pressure scale $\bP$ and mass density scale $\overline{\rho}_f$, along with the dimensionless pressure and mass density
\begin{align}
\widehat{P}(x,z) = \frac{P(x,z)}{\overline{P}},\\
\widehat{\rho}_f(x,z) = \frac{\rho_f(x,z)}{\overline{\rho}_f},
\end{align}
which are numbers of order unity. Using these dimensionless variables, Eq.~\eqref{eq:Paper3_xComponentStokes} is written as
\begin{align}
\frac{\bP}{L}\derhx{\hP}\left(\hx,\hz\right) = &\eta \left\{ \frac{\bv_x}{L^2} \derderhx{\hv_x}\left(\hx,\hz\right) + \frac{\bv_x}{h_0^2} \derderhz{\hv_x}\left(\hx,\hz\right)\right\} + \nonumber \\
                    &\left( \zeta + \frac{1}{3} \eta \right) \left\{ \frac{\bv_x}{L^2} \derderhx{\hv_x}\left(\hx,\hz\right) + \frac{\bv_z}{h_0 L} \frac{\partial^2 \hv_z}{\partial \widehat{x} \partial \widehat{z}}\left(\hx,\hz\right)\right\} - \nonumber \\
                    & - \overline{\rho}_f \widehat{\rho}_f(\hx, \hz) \left\{ \frac{ \bv_x^2}{L} \hv_x (\hx, \hz) \frac{\partial \hv_x}{\partial \hx} (\hx,\hz) \right. + \nonumber \\
                    & \left. + \frac{ \bv_z \bv_x}{h_0} \hv_z (\hx, \hz) \frac{\partial \hv_x}{\partial \hz} (\hx,\hz) \right\}, \label{Paper3_StokesGradPressureX}
\end{align}
and Eq.~\eqref{eq:Paper3_yComponentStokes} as
\begin{align}
\frac{\bP}{h_0}\derhz{\hP}\left(\hx,\hz\right) = &\eta \left\{ \frac{\bv_z}{L^2} \derderhx{\hv_z}\left(\hx,\hz\right) + \frac{\bv_z}{h_0^2} \derderhz{\hv_z}\left(\hx,\hz\right)\right\} + \nonumber \\
                    &\left( \zeta + \frac{1}{3} \eta \right) \left\{ \frac{\bv_x}{h_0L} \frac{\partial^2 \hv_x}{\partial \widehat{z} \partial \widehat{x}}\left(\hx,\hz\right)  + \frac{\bv_z}{h_0^2} \derderhz{\hv_z}\left(\hx,\hz\right)\right\} - \nonumber \\
                    &- \overline{\rho}_f \widehat{\rho}_f(\hx, \hz) \left\{ \frac{ \bv_x \bv_z}{L} \hv_x (\hx, \hz) \frac{\partial \hv_z}{\partial \hx} (\hx,\hz) \right. + \nonumber \\
                    & \left. + \frac{ \bv_z^2 }{h_0} \hv_z (\hx, \hz) \frac{\partial \hv_z}{\partial \hz} (\hx,\hz) \right\}.  \label{Paper3_StokesGradPressureY}
\end{align} 
We can relate the two velocity scales $\bv_x$ and $\bv_z$ using the continuity equation (Eq.~\eqref{eq:Paper3_continuity}), which is rewritten as
\begin{equation}
\frac{\overline{P} \overline{v}_x}{L} \frac{\partial}{\partial {\widehat{x}}} \Big[ \widehat{P}(\widehat{x}, \widehat{z}) \widehat{v}_x(\widehat{x}, \widehat{z}) \Big] + \frac{\overline{P} \overline{v}_z}{h_0} \frac{\partial}{\partial {\widehat{z}}} \Big[ \widehat{P}(\widehat{x}, \widehat{z}) \widehat{v}_z(\widehat{x}, \widehat{z}) \Big] = 0. \label{eq:Paper3_velscalesrelation0}
\end{equation}
Equation \eqref{eq:Paper3_velscalesrelation0} then yields 
\begin{equation}
\frac{\overline{v}_z}{\overline{v}_x} = -\frac{h_0}{L} \widehat{\mathcal{K}}, \label{eq:Paper3_velscalesrelation}
\end{equation}
where the non-dimensional number $\widehat{\mathcal{K}}$ is defined as 
\begin{equation}
\widehat{\mathcal{K}} =  \frac{\partial}{\partial {\widehat{x}}} \Big[ \widehat{P}(\widehat{x}, \widehat{z}) \widehat{v}_x(\widehat{x}, \widehat{z}) \Big]\left\{\frac{\partial}{\partial {\widehat{z}}} \Big[ \widehat{P}(\widehat{x}, \widehat{z}) \widehat{v}_z(\widehat{x}, \widehat{z}) \Big]\right\}^{-1}.
\end{equation}
Note that if $\frac{\partial}{\partial {\widehat{z}}} \Big[ \widehat{P}(\widehat{x}, \widehat{z}) \widehat{v}_z(\widehat{x}, \widehat{z}) \Big] = 0$, then according to Eq.~\eqref{eq:Paper3_velscalesrelation0}, so is $\frac{\partial}{\partial {\widehat{x}}} \Big[ \widehat{P}(\widehat{x}, \widehat{z}) \widehat{v}_z(\widehat{x}, \widehat{z}) \Big] = 0$, and Eq.~\eqref{eq:Paper3_velscalesrelation} still holds. Because the dimensionless velocities and positions are of order unity, so is $\widehat{\mathcal{K}}$. Plugging Eq.~\eqref{eq:Paper3_velscalesrelation} in Eq.~\eqref{Paper3_StokesGradPressureX} yields 
\begin{align}
&\frac{\bP}{L}\derhx{\hP}\left(\hx,\hz\right) = \eta \left\{ \frac{\bv_x}{L^2} \derderhx{\hv_x}\left(\hx,\hz\right) + \frac{\bv_x}{h_0^2} \derderhz{\hv_x}\left(\hx,\hz\right)\right\} + \nonumber \\
                    & + \left( \zeta + \frac{1}{3} \eta \right) \left\{ \frac{\bv_x}{L^2} \derderhx{\hv_x}\left(\hx,\hz\right) -   \frac{\bv_x}{L^2}  \widehat{\mathcal{K}}  \frac{ \partial^2\hv_z}{\partial \hx \partial \hz}\left(\hx,\hz\right)\right\} - \nonumber \\
                    & - \overline{\rho}_f \widehat{\rho}_f(\hx, \hz) \left\{ \frac{\bv_x^2}{L} \hv_x (\hx, \hz) \frac{\partial \hv_x}{\partial \hx} (\hx,\hz)  - \right. \nonumber \\
                    & \left. -  \frac{h_0}{L}\widehat{\mathcal{K}}  \frac{ \bv_x^2 }{h_0} \hv_z (\hx, \hz) \frac{\partial \hv_x}{\partial \hz} (\hx,\hz) \right\}, \label{eq:Paper3_StokesGradPressureX1}
\end{align}
and in Eq.~\eqref{Paper3_StokesGradPressureY} yields
\begin{align}
&\frac{\bP}{h_0}\derhz{\hP}\left(\hx,\hz\right) = -\eta \left\{ \frac{h_0\bv_x}{L^3} \widehat{\mathcal{K}} \derderhx{\hv_z}\left(\hx,\hz\right) + \frac{\bv_x}{h_0 L} \widehat{\mathcal{K}} \derderhz{\hv_z}\left(\hx,\hz\right)\right\} + \nonumber \\
                    & + \left( \zeta + \frac{1}{3} \eta \right) \left\{ \frac{\bv_x}{h_0L} \frac{\partial^2 \hv_x}{\partial \hz \partial \hx}\left(\hx,\hz\right)  - \frac{\bv_x}{h_0 L} \widehat{\mathcal{K}} \derderhz{\hv_z}\left(\hx,\hz\right)\right\} - \nonumber \\
                    & - \overline{\rho}_f \widehat{\rho}_f(\hx, \hz) \left\{  -\frac{h_0}{L}\widehat{\mathcal{K}}\frac{\bv_x^2}{L} \hv_x (\hx, \hz) \frac{\partial \hv_z}{\partial \hx} (\hx,\hz)  + \right. \nonumber \\
                    & \left. + \left( \frac{h_0}{L}\widehat{\mathcal{K}} \right)^2 \frac{ \bv_x^2 }{h_0} \hv_z (\hx, \hz) \frac{\partial \hv_z}{\partial \hz} (\hx,\hz) \right\},  \label{eq:Paper3_StokesGradPressureY1}
\end{align}
We now show that $\derhz{\hP}(\hx, \hz)$ is much smaller than $\derhx{\hP}(\hx, \hz)$. To do so, we calculate
\begin{align}
&\left[\derhx{\hP}\left(\hx,\hz\right)\right]^{-1} = \frac{\bP h_0^2}{L \eta \bv_x} \left\{ \left[ \derderhz{\hv_x}\left(\hx,\hz\right) + \left(\frac{h_0}{L}\right)^2 \derderhx{\hv_x}\left(\hx,\hz\right)\right] + \nonumber \right. \\
      & + \left( \frac{\zeta}{\eta} + \frac{1}{3} \right) \left[ \left(\frac{h_0}{L}\right)^2  \derderhx{\hv_x}\left(\hx,\hz\right) - \left(\frac{h_0}{L}\right)^2  \widehat{\mathcal{K}} \frac{ \partial^2\hv_z}{\partial \hx \partial \hz}\left(\hx,\hz\right)\right] -  \nonumber  \\
      &  \left. - \frac{h_0}{L} \frac{\overline{\rho}_f \bv_x h_0}{\eta} \widehat{\rho}_f(\hx,\hz)\left[ \hv_x (\hx, \hz) \frac{\partial \hv_x}{\partial \hx} (\hx,\hz) - \right. \right. \nonumber \\
      & \left. \left. - \widehat{\mathcal{K}} \hv_z (\hx, \hz) \frac{\partial \hv_x}{\partial \hz} (\hx,\hz) \right] \right\}^{-1}, \label{eq:Paper3_StokesGradPressureX3}
\end{align}
which expanded in powers of the number $h_0/L$ (which is much smaller than one), yields
\begin{align}
\left[\derhx{\hP}\left(\hx,\hz\right)\right]^{-1} = &\frac{\bP h_0^2}{L \eta \bv_x} \left[ \derderhz{\hv_x}\left(\hx,\hz\right) \right]^{-1} + \nonumber \\
& + \mathcal{O}\left[\left( \frac{h_0}{L} \right)^3 \right] + \mathcal{O} \left[ \left( \frac{h_0}{L} \right)^2 Re  \right], \label{eq:Paper3_StokesGradPressureX2}
\end{align}
where we define the Reynolds number $Re$ as 
\begin{equation}
    Re = \frac{\overline{\rho}_f \bv_x h_0}{\eta},
\end{equation}
which we take to be small such that
\begin{equation}
    \frac{h_0}{L} Re \ll 1
\end{equation}
Note that  $\derderhz{\hv_x}(\hx, \hz)$ is the largest contribution inside the curly brackets of Eq.~\eqref{eq:Paper3_StokesGradPressureX3} only in regions of space away from the points where $\derderhz{\hv_x}(\hx, \hz) =0$. While such points cannot be ruled out a priori for generic flow fields (even assuming thin reactors), they are not expected to occur often. Combining Eqs.~\eqref{eq:Paper3_StokesGradPressureX2} and \eqref{eq:Paper3_StokesGradPressureY1}, we obtain 
\begin{align}
&\derhz{\hP}\left(\hx,\hz\right)\left[\derhx{\hP}\left(\hx,\hz\right)\right]^{-1} = -\left( \frac{h_0}{L} \right)^2 \left[\derderhz{\hv_x}\left(\hx,\hz\right)\right]^{-1} \times  \nonumber \\
  & \times \left\{ \left[ \left(\frac{h_0}{L}\right)^2 \widehat{\mathcal{K}} \derderhx{\hv_z}\left(\hx,\hz\right) +  \widehat{\mathcal{K}} \derderhz{\hv_z}\left(\hx,\hz\right)\right] + \nonumber \right.\\
                    & + \left. \left( \frac{\zeta}{\eta} + \frac{1}{3} \right) \left[  \frac{\partial^2 \hv_x}{\partial \hz \partial \hx}\left(\hx,\hz\right)  - \widehat{\mathcal{K}}\derderhz{\hv_z}\left(\hx,\hz\right)\right] \vphantom{\left(\frac{h_0}{L}\right)^2} \right\} + \nonumber \\
                    & \mathcal{O} \left[ \left( \frac{h_0}{L} \right)^3\right]  + \mathcal{O} \left[ \left( \frac{h_0}{L} \right)^2 Re \right]. \label{eq:Paper3_StokesGradPressure_ratio}
\end{align} 
Assuming that the reactor is thin enough so that
\begin{equation}
\left( \frac{h_0}{L} \right)^2 \frac{\zeta}{\eta} \ll 1,
\end{equation}
one concludes from Eq.~\eqref{eq:Paper3_StokesGradPressure_ratio} that 
\begin{align}
\derhz{\hP}\left(\hx,\hz\right)  \left[\derhx{\hP}\left(\hx,\hz\right) \right]^{-1} \ll 1.
\end{align}
Accordingly, we now perform the approximation of neglecting the gradient of pressure in the transverse direction:
\begin{equation}
P(x,z) \equiv P(x),\label{eq:Paper3_thinreactorP} 
\end{equation}
and according to Eq. \eqref{eq:Paper3_idealGas}
\begin{equation}
    \rho_f(x,z) \equiv \rho_f(x).
\end{equation}
Having determined the pressure along the transverse direction, we will now turn to the number densities $\rho_A(x,z,t)$ and $\rho_B(x,z,t)$. As the sum of the two is proportional to the pressure, it is sufficient to determine one of them. Focusing on the number density of the reactant, we wish to quantify the variation of $\rho_A(x,z,t)$ in the $z$ direction and write Eq.~\eqref{eq:Paper3_ADE_A} for the steady state ($\dot{\rho}_A(x,z,t) = 0$)
\begin{align}
0 = &- \partial_x [\rho_A(x,z) \bm{v}_x(x,z)] - \partial_z [\rho_A(x,z) \bm{v}_z(x,z)] + \nonumber \\
& +D \partial_z^2 \rho_A(x,z)  - \xi(x,z), \label{eq:Paper3_gettingrhoY}
\end{align}
where the diffusive flux in the $x$ direction has been neglected as per Eq.~\eqref{eq:Paper3_diffXsmall}. Applying the same reasoning as for the pressure, we write Eq.~\eqref{eq:Paper3_gettingrhoY} as
\begin{align}
0 = &- \frac{\overline{\rho}_A \bv_x }{L} \derhx{}\left[\hat{\rho}_A (\hx,\hz) \hv_x (\hx,\hz)\right] - \frac{\overline{\rho}_A \bv_z }{h_0} \derhz{}\left[\hat{\rho}_A (\hx,\hz) \hv_z (\hx,\hz)\right] + \nonumber \\
    & + \frac{D \overline{\rho}_A}{h_0^2} \derderhz{\hat{\rho}_A}(\hx,\hz)  - \nonumber \\
    &- \frac{\alpha(\hx) \overline{\rho}}{h_0} \left\{\delta\left[\hz-\frac{h(\hx)}{h_0}\right]+\delta\left[\hz+\frac{h(\hx)}{h_0}\right]\right\}, \label{eq:Paper3_gettingrhoY2}
\end{align}
where we have used Eq.~\eqref{eq:Paper3_xi}. The Dirac deltas in Eq.~\eqref{eq:Paper3_gettingrhoY2} are now non-dimensional, as the arguments $\hz+ h(\hx)/h_0$ are also non-dimensional. Using Eq.~\eqref{eq:Paper3_velscalesrelation}, we obtain 
\begin{align}
0 = & \frac{\overline{\rho}_A \bv_z }{h_0} \frac{1}{\widehat{\mathcal{K}}} \derhx{}\left[\hat{\rho}_A (\hx,\hz) \hv_x (\hx,\hz)\right] - \frac{\overline{\rho}_A \bv_z }{h_0} \derhz{}\left[\hat{\rho}_A (\hx,\hz) \hv_z (\hx,\hz)\right] + \nonumber \\
    & + \frac{D \overline{\rho}_A}{h_0^2} \derderhz{\hat{\rho}_A}(\hx,\hz) - \frac{\alpha(\hx) \overline{\rho}}{h_0} \left\{\delta\left[\hz-\frac{h(\hx)}{h_0}\right]+\delta\left[\hz+\frac{h(\hx)}{h_0}\right]\right\}, 
\end{align}
which can be rewritten as 
\begin{align}
 \derderhz{\hat{\rho}_A}(\hx,\hz) = & -\frac{\bv_z h_0}{D} \frac{1}{\widehat{\mathcal{K}}} \derhx{}\left[\hat{\rho}_A (\hx,\hz) \hv_x (\hx,\hz)\right] + \nonumber \\
    &  +  \frac{\bv_z h_0}{D}  \derhz{}\left[\hat{\rho}_A (\hx,\hz) \hv_z (\hx,\hz)\right] + \nonumber \\
    &  +  \frac{\alpha(\hx) h_0}{D} \left\{\delta\left[\hz-\frac{h(\hx)}{h_0}\right]+\delta\left[\hz+\frac{h(\hx)}{h_0}\right]\right\}.
\end{align}
Integrating along the $z$ direction from $-\infty$ to $\hat{z}$ yields
\begin{align}
 \derhz{\hat{\rho}_A}(\hx,\hz) = & -\frac{\bv_z h_0}{D} \frac{1}{\widehat{\mathcal{K}}} \int\limits_{-\infty}^{\hz} \derhx{}\left[\hat{\rho}_A (\hx,\hz^{\prime}) \hv_x (\hx,\hz^{\prime})\right] d\hz^{\prime} + \nonumber \\
    &  +  \frac{\bv_z h_0}{D}  \hat{\rho}_A (\hx,\hz) \hv_z (\hx,\hz)  +  \frac{\alpha(\hx) h_0}{D}.
\end{align}
To proceed, we assume the reactor to be so thin that the P\'eclet number $Pe_z$ along the $z$ direction is much smaller than one
\begin{equation}
    Pe_z = \frac{\overline{v}_z h_0}{D} \ll 1.\label{eq:Paper3_Pe_z}
\end{equation}
We also identify the Damk\"ohler number in the transverse direction $Da_z$. This number is the ratio of the timescale associated with diffusing along the cross-section $h_0^2/D$ and the timescale associated with the reaction $h_0/\overline{\alpha}$
\begin{equation}
    Da = \frac{\overline{\alpha} h_0}{D} 
\end{equation}
where $\overline{\alpha}$ is the typical scale for the reaction rate. We also assume the channel to be thin enough so that $Da_z$ is small
\begin{equation}
    Da \ll 1, \label{eq:Paper3_Da}
\end{equation}
and as a result
\begin{equation}
 \frac{1}{\hat{\rho}_A(\hx,\hz)}\derhz{\hat{\rho}_A}(\hx,\hz) \ll 1,
\end{equation}
and thus the number density of both reactant and product may be approximated as constant in the transverse direction 
\begin{align}
    \rho_A (x,z) &\equiv \rho_A(x), \label{eq:Paper3_rhoA_x} \\
    \rho_B (x,z) &\equiv \rho_B(x). \label{eq:Paper3_rhoB_x} 
\end{align}
With the approximation of Eq.~\eqref{eq:Paper3_thinreactorP}, we may rewrite Eq.~\eqref{eq:Paper3_StokesGradPressureX1} as 
\begin{align}
&\left(\frac{h_0}{L}\right)^2 \bP\derhx{\hP}\left(\hx\right) = \eta \frac{\bv_x}{L}\left\{ \left(\frac{h_0}{L}\right)^2  \derderhx{\hv_x}\left(\hx,\hz\right) +  \derderhz{\hv_x}\left(\hx,\hz\right)\right\} + \nonumber \\
                    &+ \left(\frac{h_0}{L}\right)^2  \left( \zeta + \frac{1}{3} \eta \right) \frac{\bv_x}{L}  \left\{ \derderhx{\hv_x}\left(\hx,\hz\right) + \widehat{\mathcal{K}} \frac{\partial^2\hv_z}{\partial \hx \partial \hz}\left(\hx,\hz\right)\right\} - \nonumber \\
                    & - \frac{h_0}{L}Re \eta \frac{\bv_x}{L} \widehat{\rho}_f(\hx) \left[   \hv_x (\hx,\hz) \frac{\partial \hv_x}{ \partial \hx} (\hx,\hz) - \widehat{\mathcal{K}} \hv_z (\hx, \hz) \frac{\partial \hv_x}{\partial \hz} (\hx,\hz) \right], 
\end{align}
and as $h_0/L \ll 1$, we employ the lubrication approximation \cite{Schlichting1979} and neglect all terms on the right-hand-side that are not of the order of $(h_0/L)^2$ or $Re(h_0/L)$:
\begin{equation}
\partial_x P(x)  =  \eta \partial_z^2 v_x(x,z) + \mathcal{O}\left[ \left( \frac{h_0}{L} \right)^2 \right] + \mathcal{O}\left[ \frac{h_0}{L} Re \right],
\end{equation}
where we have returned to dimensional variables for clarity of reading. Integrating both sides over the $z$ direction twice yields
\begin{equation}
 v_{x}(x,z)=\frac{z^{2}-h^{2}(x)}{2\eta}\partial_x P(x), \label{eq:Paper3_v_x} 
\end{equation}
where we have used
\begin{equation}
v_x [x, \pm h(x)] = 0, \label{eq:Paper3_noSlip}
\end{equation}
coming from the no-slip boundary condition. The flow described in Eq. \eqref{eq:Paper3_v_x} is Poiseuille-like with a local pressure gradient $\partial_x P(x)$.

\section{Experimental methodology}

\label{sec:experiments}

Reagents for the preparation of the samples were purchased from Sigma–Aldrich, Alfa Aesar, Strem, or VWR and used as received. A Millipore Direct-Q system was employed for the purification of water before use. Aluminum plates (99.99\%) for the anodization process were purchased from Smart Membranes GmbH. The generation of the ordered system of parallel nanopores involves a two-step anodization procedure described in the literature \cite{Englhard2021, Ruiz2023}. The specific anodization parameters used here include the applied voltage 195 V, the electrolyte solution 1 wt.\% \ce{H3PO4}, the anodization temperature 0\textdegree C,  and its duration 18 h. Thereafter, the remaining metallic aluminium substrate was removed by a \ce{CuCl2} solution (0.7 M in 10\% \ce{HCl}). Finally, the samples were treated with a 10\% v/v \ce{H3PO4} solution for 60 min. at 45\textdegree C to etch the \ce{Al2O3} barrier layer, open the lower pore extremities and widening the pore diameter to approximately 400 nm. These nanoporous membranes were finally laser-cut into individual nanoreactors with 4 mm diameter. Catalytically active SCALMS nanoparticles with a diameter of 30 $\pm$ 5 nm were synthesized by a hot injection method adapted from a published procedure \cite{Yarema2014}. They were suspended in chloroform (1 mg/mL) and then infiltrated into the nanoporous reactors by placing 25 $\upmu$L of suspension on a 4-mm reactor, then passing the liquid using vacuum filtration, and repeating the procedure four times in total. The sample was washed with acetone, isopropanol, and finally with distilled water, then left to dry under vacuum.  

The catalytic model reaction adapts a procedure published by Rahim et al. \cite{Rahim2022}, diluting the 100 mg/L methylene blue solution in 0.1 M aqueous \ce{HCl} further to 10 $\upmu$M (or 3.6 mg/L, a dilution by a factor 28). 100 $\upmu$L of a freshly prepared ascorbic acid solution (1 mg/mL in water) were added to a glass vial containing 30 mL of the methylene blue solution, then the mixture was flown at a rate of 100 $\upmu$L/min through the nanoreactor, consisting of a SCALMS-loaded anodized alumina membrane of 4 mm diameter held in a Kapton ring and sandwiched between both parts of a 3D-printed PMMA holder. The holder consists of two parts in Swagelok VCR geometry that seal on the Kapton ring while supporting the porous nanoreactor on a backing grid. The reaction medium was collected in a vial in individual aliquots and analyzed by UV-Vis spectrometry.

The catalyst distribution and chemical composition along the nanopores were analyzed using a JSM-F100 field-emission scanning electron microscope from JEOL equipped with EDX detector. These results were averaged over ten samples. The only nanoparticles that deposit close to the pore outlet are the ones that avoided deposition further upstream. As such, the closer to the outlet, the less likely it is for a nanoparticle to deposit, leading to a fore-aft asymmetric distribution of catalytic material inside the pore.

\bibliography{refs}

\begin{thebibliography}{78}%
\makeatletter
\providecommand \@ifxundefined [1]{%
 \@ifx{#1\undefined}
}%
\providecommand \@ifnum [1]{%
 \ifnum #1\expandafter \@firstoftwo
 \else \expandafter \@secondoftwo
 \fi
}%
\providecommand \@ifx [1]{%
 \ifx #1\expandafter \@firstoftwo
 \else \expandafter \@secondoftwo
 \fi
}%
\providecommand \natexlab [1]{#1}%
\providecommand \enquote  [1]{``#1''}%
\providecommand \bibnamefont  [1]{#1}%
\providecommand \bibfnamefont [1]{#1}%
\providecommand \citenamefont [1]{#1}%
\providecommand \href@noop [0]{\@secondoftwo}%
\providecommand \href [0]{\begingroup \@sanitize@url \@href}%
\providecommand \@href[1]{\@@startlink{#1}\@@href}%
\providecommand \@@href[1]{\endgroup#1\@@endlink}%
\providecommand \@sanitize@url [0]{\catcode `\\12\catcode `\$12\catcode
  `\&12\catcode `\#12\catcode `\^12\catcode `\_12\catcode `\%12\relax}%
\providecommand \@@startlink[1]{}%
\providecommand \@@endlink[0]{}%
\providecommand \url  [0]{\begingroup\@sanitize@url \@url }%
\providecommand \@url [1]{\endgroup\@href {#1}{\urlprefix }}%
\providecommand \urlprefix  [0]{URL }%
\providecommand \Eprint [0]{\href }%
\providecommand \doibase [0]{http://dx.doi.org/}%
\providecommand \selectlanguage [0]{\@gobble}%
\providecommand \bibinfo  [0]{\@secondoftwo}%
\providecommand \bibfield  [0]{\@secondoftwo}%
\providecommand \translation [1]{[#1]}%
\providecommand \BibitemOpen [0]{}%
\providecommand \bibitemStop [0]{}%
\providecommand \bibitemNoStop [0]{.\EOS\space}%
\providecommand \EOS [0]{\spacefactor3000\relax}%
\providecommand \BibitemShut  [1]{\csname bibitem#1\endcsname}%
\let\auto@bib@innerbib\@empty
\bibitem [{\citenamefont {Hagen}(2015)}]{book_IndustrialCatalysis}%
  \BibitemOpen
  \bibfield  {author} {\bibinfo {author} {\bibfnamefont {J.}~\bibnamefont
  {Hagen}},\ }\href@noop {} {\emph {\bibinfo {title} {Industrial Catalysis: a
  Practical Approach}}}\ (\bibinfo  {publisher} {Wiley-VCH},\ \bibinfo {year}
  {2015})\BibitemShut {NoStop}%
\bibitem [{\citenamefont {Tanimu}\ \emph {et~al.}(2022)\citenamefont {Tanimu},
  \citenamefont {Tanimu}, \citenamefont {Alasiri},\ and\ \citenamefont
  {Aitani}}]{Tanimu2022}%
  \BibitemOpen
  \bibfield  {author} {\bibinfo {author} {\bibfnamefont {A.}~\bibnamefont
  {Tanimu}}, \bibinfo {author} {\bibfnamefont {G.}~\bibnamefont {Tanimu}},
  \bibinfo {author} {\bibfnamefont {H.}~\bibnamefont {Alasiri}}, \ and\
  \bibinfo {author} {\bibfnamefont {A.}~\bibnamefont {Aitani}},\ }\bibfield
  {title} {\enquote {\bibinfo {title} {Catalytic cracking of crude oil: Mini
  review of catalyst formulations for enhanced selectivity to light olefins},}\
  }\href@noop {} {\bibfield  {journal} {\bibinfo  {journal} {Energy Fuels}\
  }\textbf {\bibinfo {volume} {36}},\ \bibinfo {pages} {5152--5166} (\bibinfo
  {year} {2022})}\BibitemShut {NoStop}%
\bibitem [{\citenamefont {Zhu}, \citenamefont {Liang},\ and\ \citenamefont
  {Zou}(2020)}]{Zhu2020}%
  \BibitemOpen
  \bibfield  {author} {\bibinfo {author} {\bibfnamefont {B.}~\bibnamefont
  {Zhu}}, \bibinfo {author} {\bibfnamefont {Z.}~\bibnamefont {Liang}}, \ and\
  \bibinfo {author} {\bibfnamefont {R.}~\bibnamefont {Zou}},\ }\bibfield
  {title} {\enquote {\bibinfo {title} {Designing advanced catalysts for energy
  conversion based on urea oxidation reaction},}\ }\href@noop {} {\bibfield
  {journal} {\bibinfo  {journal} {Small}\ }\textbf {\bibinfo {volume} {16}},\
  \bibinfo {pages} {1906133} (\bibinfo {year} {2020})}\BibitemShut {NoStop}%
\bibitem [{\citenamefont {Pei}\ \emph {et~al.}(2018)\citenamefont {Pei},
  \citenamefont {Cheng}, \citenamefont {Chen}, \citenamefont {Smith},
  \citenamefont {Dong}, \citenamefont {Ajayan}, \citenamefont {Ye},\ and\
  \citenamefont {Shen}}]{Pei2018}%
  \BibitemOpen
  \bibfield  {author} {\bibinfo {author} {\bibfnamefont {Y.}~\bibnamefont
  {Pei}}, \bibinfo {author} {\bibfnamefont {Y.}~\bibnamefont {Cheng}}, \bibinfo
  {author} {\bibfnamefont {J.}~\bibnamefont {Chen}}, \bibinfo {author}
  {\bibfnamefont {W.}~\bibnamefont {Smith}}, \bibinfo {author} {\bibfnamefont
  {P.}~\bibnamefont {Dong}}, \bibinfo {author} {\bibfnamefont {P.~M.}\
  \bibnamefont {Ajayan}}, \bibinfo {author} {\bibfnamefont {M.}~\bibnamefont
  {Ye}}, \ and\ \bibinfo {author} {\bibfnamefont {J.}~\bibnamefont {Shen}},\
  }\bibfield  {title} {\enquote {\bibinfo {title} {Recent developments of
  transition metal phosphides as catalysts in the energy conversion field},}\
  }\href@noop {} {\bibfield  {journal} {\bibinfo  {journal} {J. Mater. Chem.
  A}\ }\textbf {\bibinfo {volume} {6}},\ \bibinfo {pages} {23220--23243}
  (\bibinfo {year} {2018})}\BibitemShut {NoStop}%
\bibitem [{\citenamefont {Shekhawat}, \citenamefont {Spivery},\ and\
  \citenamefont {Berry}(2011)}]{book_FuelCells}%
  \BibitemOpen
  \bibfield  {author} {\bibinfo {author} {\bibfnamefont {D.}~\bibnamefont
  {Shekhawat}}, \bibinfo {author} {\bibfnamefont {J.~J.}\ \bibnamefont
  {Spivery}}, \ and\ \bibinfo {author} {\bibfnamefont {D.}~\bibnamefont
  {Berry}},\ }\href@noop {} {\emph {\bibinfo {title} {Fuel Cells: Technologies
  for Fuel Processing}}}\ (\bibinfo  {publisher} {Elsevier Science},\ \bibinfo
  {year} {2011})\BibitemShut {NoStop}%
\bibitem [{\citenamefont {Anastas}, \citenamefont {Kirchhoff},\ and\
  \citenamefont {Williamson}(2001)}]{Anastas2001}%
  \BibitemOpen
  \bibfield  {author} {\bibinfo {author} {\bibfnamefont {P.~T.}\ \bibnamefont
  {Anastas}}, \bibinfo {author} {\bibfnamefont {M.~M.}\ \bibnamefont
  {Kirchhoff}}, \ and\ \bibinfo {author} {\bibfnamefont {T.~C.}\ \bibnamefont
  {Williamson}},\ }\bibfield  {title} {\enquote {\bibinfo {title} {Catalysis as
  a foundational pillar of green chemistry},}\ }\href@noop {} {\bibfield
  {journal} {\bibinfo  {journal} {Appl. Catal. A: Gen}\ }\textbf {\bibinfo
  {volume} {221}},\ \bibinfo {pages} {3--13} (\bibinfo {year}
  {2001})}\BibitemShut {NoStop}%
\bibitem [{\citenamefont {Anastas}\ and\ \citenamefont
  {Kirchhoff}(2002)}]{Anastas2002}%
  \BibitemOpen
  \bibfield  {author} {\bibinfo {author} {\bibfnamefont {P.~T.}\ \bibnamefont
  {Anastas}}\ and\ \bibinfo {author} {\bibfnamefont {M.~M.}\ \bibnamefont
  {Kirchhoff}},\ }\bibfield  {title} {\enquote {\bibinfo {title} {Origins,
  current status, and future challenges of green chemistry},}\ }\href@noop {}
  {\bibfield  {journal} {\bibinfo  {journal} {Acc. Chem. Res.}\ }\textbf
  {\bibinfo {volume} {35}},\ \bibinfo {pages} {686--694} (\bibinfo {year}
  {2002})}\BibitemShut {NoStop}%
\bibitem [{\citenamefont {Zimmerman}\ \emph {et~al.}(2020)\citenamefont
  {Zimmerman}, \citenamefont {Anastas}, \citenamefont {Erythropel},\ and\
  \citenamefont {Leitner}}]{Zimmerman2020}%
  \BibitemOpen
  \bibfield  {author} {\bibinfo {author} {\bibfnamefont {J.~B.}\ \bibnamefont
  {Zimmerman}}, \bibinfo {author} {\bibfnamefont {P.~T.}\ \bibnamefont
  {Anastas}}, \bibinfo {author} {\bibfnamefont {H.~C.}\ \bibnamefont
  {Erythropel}}, \ and\ \bibinfo {author} {\bibfnamefont {W.}~\bibnamefont
  {Leitner}},\ }\bibfield  {title} {\enquote {\bibinfo {title} {Designing for a
  green chemistry future},}\ }\href@noop {} {\bibfield  {journal} {\bibinfo
  {journal} {Science}\ }\textbf {\bibinfo {volume} {367}},\ \bibinfo {pages}
  {397--400} (\bibinfo {year} {2020})}\BibitemShut {NoStop}%
\bibitem [{\citenamefont {Riisager}\ \emph {et~al.}(2006)\citenamefont
  {Riisager}, \citenamefont {Fehrmann}, \citenamefont {Haumann},\ and\
  \citenamefont {Wasserscheid}}]{Riisager2006}%
  \BibitemOpen
  \bibfield  {author} {\bibinfo {author} {\bibfnamefont {A.}~\bibnamefont
  {Riisager}}, \bibinfo {author} {\bibfnamefont {R.}~\bibnamefont {Fehrmann}},
  \bibinfo {author} {\bibfnamefont {M.}~\bibnamefont {Haumann}}, \ and\
  \bibinfo {author} {\bibfnamefont {P.}~\bibnamefont {Wasserscheid}},\
  }\bibfield  {title} {\enquote {\bibinfo {title} {Supported ionic liquid phase
  (silp) catalysis: An innovative concept for homogeneous catalysis in
  continuous fixed-bed reactors},}\ }\href@noop {} {\bibfield  {journal}
  {\bibinfo  {journal} {Eur. J. Inorg. Chem.}\ }\textbf {\bibinfo {volume}
  {2006}},\ \bibinfo {pages} {695--706} (\bibinfo {year} {2006})}\BibitemShut
  {NoStop}%
\bibitem [{\citenamefont {Riisager}\ \emph {et~al.}(2008)\citenamefont
  {Riisager}, \citenamefont {Fehrmann}, \citenamefont {Haumann},\ and\
  \citenamefont {Wasserscheid}}]{Riisager2008}%
  \BibitemOpen
  \bibfield  {author} {\bibinfo {author} {\bibfnamefont {A.}~\bibnamefont
  {Riisager}}, \bibinfo {author} {\bibfnamefont {R.}~\bibnamefont {Fehrmann}},
  \bibinfo {author} {\bibfnamefont {M.}~\bibnamefont {Haumann}}, \ and\
  \bibinfo {author} {\bibfnamefont {P.}~\bibnamefont {Wasserscheid}},\
  }\enquote {\bibinfo {title} {Catalytic silp materials},}\ in\ \href@noop {}
  {\emph {\bibinfo {booktitle} {Regulated Systems for Multiphase Catalysis}}},\
  \bibinfo {editor} {edited by\ \bibinfo {editor} {\bibfnamefont
  {W.}~\bibnamefont {Leitner}}\ and\ \bibinfo {editor} {\bibfnamefont
  {M.}~\bibnamefont {H{\"o}lscher}}}\ (\bibinfo  {publisher} {Springer Berlin
  Heidelberg},\ \bibinfo {address} {Berlin, Heidelberg},\ \bibinfo {year}
  {2008})\ pp.\ \bibinfo {pages} {149--161}\BibitemShut {NoStop}%
\bibitem [{\citenamefont {Taccardi}\ \emph {et~al.}(2017)\citenamefont
  {Taccardi}, \citenamefont {Grabau}, \citenamefont {Debuschewitz},
  \citenamefont {Distaso}, \citenamefont {Brandl}, \citenamefont {Hock},
  \citenamefont {Maier}, \citenamefont {Papp}, \citenamefont {Erhard},
  \citenamefont {Neiss}, \citenamefont {Peukert}, \citenamefont {G{\"o}rling},
  \citenamefont {Steinr{\"u}ck},\ and\ \citenamefont
  {Wasserscheid}}]{Taccardi2017}%
  \BibitemOpen
  \bibfield  {author} {\bibinfo {author} {\bibfnamefont {N.}~\bibnamefont
  {Taccardi}}, \bibinfo {author} {\bibfnamefont {M.}~\bibnamefont {Grabau}},
  \bibinfo {author} {\bibfnamefont {J.}~\bibnamefont {Debuschewitz}}, \bibinfo
  {author} {\bibfnamefont {M.}~\bibnamefont {Distaso}}, \bibinfo {author}
  {\bibfnamefont {M.}~\bibnamefont {Brandl}}, \bibinfo {author} {\bibfnamefont
  {R.}~\bibnamefont {Hock}}, \bibinfo {author} {\bibfnamefont {F.}~\bibnamefont
  {Maier}}, \bibinfo {author} {\bibfnamefont {C.}~\bibnamefont {Papp}},
  \bibinfo {author} {\bibfnamefont {J.}~\bibnamefont {Erhard}}, \bibinfo
  {author} {\bibfnamefont {C.}~\bibnamefont {Neiss}}, \bibinfo {author}
  {\bibfnamefont {W.}~\bibnamefont {Peukert}}, \bibinfo {author} {\bibfnamefont
  {A.}~\bibnamefont {G{\"o}rling}}, \bibinfo {author} {\bibfnamefont {H.-P.}\
  \bibnamefont {Steinr{\"u}ck}}, \ and\ \bibinfo {author} {\bibfnamefont
  {P.}~\bibnamefont {Wasserscheid}},\ }\bibfield  {title} {\enquote {\bibinfo
  {title} {Gallium-rich pd--ga phases as supported liquid metal catalysts},}\
  }\href {\doibase 10.1038/nchem.2822} {\bibfield  {journal} {\bibinfo
  {journal} {Nat. Chem.}\ }\textbf {\bibinfo {volume} {9}},\ \bibinfo {pages}
  {862--867} (\bibinfo {year} {2017})}\BibitemShut {NoStop}%
\bibitem [{\citenamefont {Rupprechter}(2017)}]{Rupprechter2017}%
  \BibitemOpen
  \bibfield  {author} {\bibinfo {author} {\bibfnamefont {G.}~\bibnamefont
  {Rupprechter}},\ }\bibfield  {title} {\enquote {\bibinfo {title} {Popping up
  to the surface},}\ }\href {\doibase 10.1038/nchem.2849} {\bibfield  {journal}
  {\bibinfo  {journal} {Nat. Chem.}\ }\textbf {\bibinfo {volume} {9}},\
  \bibinfo {pages} {833--834} (\bibinfo {year} {2017})}\BibitemShut {NoStop}%
\bibitem [{\citenamefont {Kernchen}\ \emph {et~al.}(2007)\citenamefont
  {Kernchen}, \citenamefont {Etzold}, \citenamefont {Korth},\ and\
  \citenamefont {Jess}}]{Kernchen2007}%
  \BibitemOpen
  \bibfield  {author} {\bibinfo {author} {\bibfnamefont {U.}~\bibnamefont
  {Kernchen}}, \bibinfo {author} {\bibfnamefont {B.}~\bibnamefont {Etzold}},
  \bibinfo {author} {\bibfnamefont {W.}~\bibnamefont {Korth}}, \ and\ \bibinfo
  {author} {\bibfnamefont {A.}~\bibnamefont {Jess}},\ }\bibfield  {title}
  {\enquote {\bibinfo {title} {Solid catalyst with ionic liquid layer (scill)
  – a new concept to improve selectivity illustrated by hydrogenation of
  cyclooctadiene},}\ }\href {\doibase https://doi.org/10.1002/ceat.200700050}
  {\bibfield  {journal} {\bibinfo  {journal} {Chemical Engineering \&
  Technology}\ }\textbf {\bibinfo {volume} {30}},\ \bibinfo {pages} {985--994}
  (\bibinfo {year} {2007})},\ \Eprint
  {http://arxiv.org/abs/https://onlinelibrary.wiley.com/doi/pdf/10.1002/ceat.200700050}
  {https://onlinelibrary.wiley.com/doi/pdf/10.1002/ceat.200700050} \BibitemShut
  {NoStop}%
\bibitem [{\citenamefont {Boger}, \citenamefont {Heibel},\ and\ \citenamefont
  {Sorensen}(2004)}]{Boger2004}%
  \BibitemOpen
  \bibfield  {author} {\bibinfo {author} {\bibfnamefont {T.}~\bibnamefont
  {Boger}}, \bibinfo {author} {\bibfnamefont {A.~K.}\ \bibnamefont {Heibel}}, \
  and\ \bibinfo {author} {\bibfnamefont {C.~M.}\ \bibnamefont {Sorensen}},\
  }\bibfield  {title} {\enquote {\bibinfo {title} {Monolithic catalysts for the
  chemical industry},}\ }\href@noop {} {\bibfield  {journal} {\bibinfo
  {journal} {Ind. Eng. Chem. Res.}\ }\textbf {\bibinfo {volume} {43}},\
  \bibinfo {pages} {4602--4611} (\bibinfo {year} {2004})}\BibitemShut {NoStop}%
\bibitem [{\citenamefont {\"Onsen}\ and\ \citenamefont
  {Avci}(2016)}]{Onsen2016}%
  \BibitemOpen
  \bibfield  {author} {\bibinfo {author} {\bibfnamefont {Z.~I.}\ \bibnamefont
  {\"Onsen}}\ and\ \bibinfo {author} {\bibfnamefont {A.~K.}\ \bibnamefont
  {Avci}},\ }\href@noop {} {\emph {\bibinfo {title} {Multiphase Catalytic
  Reactors: Theory, Design, Manufacturing, and Applications}}}\ (\bibinfo
  {publisher} {John Wiley \& Sons, Ltd},\ \bibinfo {year} {2016})\ pp.\
  \bibinfo {pages} {171--212}\BibitemShut {NoStop}%
\bibitem [{\citenamefont {{du Plessis}}\ and\ \citenamefont
  {Woudberg}(2008)}]{Woudberg2008}%
  \BibitemOpen
  \bibfield  {author} {\bibinfo {author} {\bibfnamefont {J.~P.}\ \bibnamefont
  {{du Plessis}}}\ and\ \bibinfo {author} {\bibnamefont {Woudberg}},\
  }\bibfield  {title} {\enquote {\bibinfo {title} {Pore-scale derivation of the
  ergun equation to enhance its adaptability and generalization},}\ }\href@noop
  {} {\bibfield  {journal} {\bibinfo  {journal} {Chemical Engineering Science}\
  }\textbf {\bibinfo {volume} {63}},\ \bibinfo {pages} {2576--2586} (\bibinfo
  {year} {2008})}\BibitemShut {NoStop}%
\bibitem [{\citenamefont {Macdonald}\ \emph {et~al.}(1979)\citenamefont
  {Macdonald}, \citenamefont {El-Sayed}, \citenamefont {Mow},\ and\
  \citenamefont {Dullien}}]{Macdonald1979}%
  \BibitemOpen
  \bibfield  {author} {\bibinfo {author} {\bibfnamefont {I.~F.}\ \bibnamefont
  {Macdonald}}, \bibinfo {author} {\bibfnamefont {M.~S.}\ \bibnamefont
  {El-Sayed}}, \bibinfo {author} {\bibfnamefont {K.}~\bibnamefont {Mow}}, \
  and\ \bibinfo {author} {\bibfnamefont {F.~A.~L.}\ \bibnamefont {Dullien}},\
  }\bibfield  {title} {\enquote {\bibinfo {title} {Flow through porous
  media-the ergun equation revisited},}\ }\href@noop {} {\bibfield  {journal}
  {\bibinfo  {journal} {Industrial {\&} Engineering Chemistry Fundamentals}\
  }\textbf {\bibinfo {volume} {18}},\ \bibinfo {pages} {199--208} (\bibinfo
  {year} {1979})}\BibitemShut {NoStop}%
\bibitem [{\citenamefont {Bachmat}\ and\ \citenamefont
  {Bear}(1986)}]{Bachmat1986}%
  \BibitemOpen
  \bibfield  {author} {\bibinfo {author} {\bibfnamefont {Y.}~\bibnamefont
  {Bachmat}}\ and\ \bibinfo {author} {\bibfnamefont {J.}~\bibnamefont {Bear}},\
  }\bibfield  {title} {\enquote {\bibinfo {title} {Macroscopic modelling of
  transport phenomena in porous media. 1: The continuum approach},}\
  }\href@noop {} {\bibfield  {journal} {\bibinfo  {journal} {Transport in
  Porous Media}\ }\textbf {\bibinfo {volume} {1}},\ \bibinfo {pages} {213--240}
  (\bibinfo {year} {1986})}\BibitemShut {NoStop}%
\bibitem [{\citenamefont {Bear}\ and\ \citenamefont
  {Bachmat}(1986)}]{Bear1986}%
  \BibitemOpen
  \bibfield  {author} {\bibinfo {author} {\bibfnamefont {J.}~\bibnamefont
  {Bear}}\ and\ \bibinfo {author} {\bibfnamefont {Y.}~\bibnamefont {Bachmat}},\
  }\bibfield  {title} {\enquote {\bibinfo {title} {Macroscopic modelling of
  transport phenomena in porous media. 2: Applications to mass, momentum and
  energy transport},}\ }\href@noop {} {\bibfield  {journal} {\bibinfo
  {journal} {Transport in Porous Media}\ }\textbf {\bibinfo {volume} {1}},\
  \bibinfo {pages} {241--269} (\bibinfo {year} {1986})}\BibitemShut {NoStop}%
\bibitem [{\citenamefont {Negrini}\ \emph {et~al.}(1999)\citenamefont
  {Negrini}, \citenamefont {Fuelber}, \citenamefont {Freire},\ and\
  \citenamefont {Thom\'eo}}]{Negrini1999}%
  \BibitemOpen
  \bibfield  {author} {\bibinfo {author} {\bibfnamefont {A.~L.}\ \bibnamefont
  {Negrini}}, \bibinfo {author} {\bibfnamefont {A.}~\bibnamefont {Fuelber}},
  \bibinfo {author} {\bibfnamefont {J.~T.}\ \bibnamefont {Freire}}, \ and\
  \bibinfo {author} {\bibfnamefont {J.~C.}\ \bibnamefont {Thom\'eo}},\
  }\bibfield  {title} {\enquote {\bibinfo {title} {Fluid dynamics of air in a
  packed bed: velocity profiles and the continuum model assumption},}\
  }\href@noop {} {\bibfield  {journal} {\bibinfo  {journal} {Braz. J. Chem.}\
  }\textbf {\bibinfo {volume} {16}} (\bibinfo {year} {1999})}\BibitemShut
  {NoStop}%
\bibitem [{\citenamefont {Battiato}\ and\ \citenamefont
  {Tartakovsky}(2011)}]{Battiato2011}%
  \BibitemOpen
  \bibfield  {author} {\bibinfo {author} {\bibfnamefont {I.}~\bibnamefont
  {Battiato}}\ and\ \bibinfo {author} {\bibfnamefont {D.~M.}\ \bibnamefont
  {Tartakovsky}},\ }\bibfield  {title} {\enquote {\bibinfo {title}
  {Applicability regimes for macroscopic models of reactive transport in porous
  media},}\ }\href@noop {} {\bibfield  {journal} {\bibinfo  {journal} {Journal
  of Contaminant Hydrology}\ }\textbf {\bibinfo {volume} {120-121}},\ \bibinfo
  {pages} {18--26} (\bibinfo {year} {2011})}\BibitemShut {NoStop}%
\bibitem [{\citenamefont {Heck}, \citenamefont {Farrauto},\ and\ \citenamefont
  {Gulati}(2009)}]{book_catalyticAirPollution}%
  \BibitemOpen
  \bibfield  {author} {\bibinfo {author} {\bibfnamefont {R.~M.}\ \bibnamefont
  {Heck}}, \bibinfo {author} {\bibfnamefont {R.~J.}\ \bibnamefont {Farrauto}},
  \ and\ \bibinfo {author} {\bibfnamefont {S.~T.}\ \bibnamefont {Gulati}},\
  }\href@noop {} {\emph {\bibinfo {title} {Catalytic Air Pollution Control:
  Commercial Technology}}}\ (\bibinfo  {publisher} {Wiley},\ \bibinfo {year}
  {2009})\BibitemShut {NoStop}%
\bibitem [{\citenamefont {Ran}\ \emph {et~al.}(2023)\citenamefont {Ran},
  \citenamefont {Wang}, \citenamefont {Liu}, \citenamefont {Yin}, \citenamefont
  {Li},\ and\ \citenamefont {Zhang}}]{Ran2023}%
  \BibitemOpen
  \bibfield  {author} {\bibinfo {author} {\bibfnamefont {J.}~\bibnamefont
  {Ran}}, \bibinfo {author} {\bibfnamefont {X.}~\bibnamefont {Wang}}, \bibinfo
  {author} {\bibfnamefont {Y.}~\bibnamefont {Liu}}, \bibinfo {author}
  {\bibfnamefont {S.}~\bibnamefont {Yin}}, \bibinfo {author} {\bibfnamefont
  {S.}~\bibnamefont {Li}}, \ and\ \bibinfo {author} {\bibfnamefont
  {L.}~\bibnamefont {Zhang}},\ }\bibfield  {title} {\enquote {\bibinfo {title}
  {Microreactor-based micro/nanomaterials: fabrication{,} advances{,} and
  outlook},}\ }\href {\doibase 10.1039/D3MH00329A} {\bibfield  {journal}
  {\bibinfo  {journal} {Mater. Horiz.}\ }\textbf {\bibinfo {volume} {10}},\
  \bibinfo {pages} {2343--2372} (\bibinfo {year} {2023})}\BibitemShut {NoStop}%
\bibitem [{\citenamefont {Kolb}\ and\ \citenamefont {Hessel}(2004)}]{Kolb2004}%
  \BibitemOpen
  \bibfield  {author} {\bibinfo {author} {\bibfnamefont {G.}~\bibnamefont
  {Kolb}}\ and\ \bibinfo {author} {\bibfnamefont {V.}~\bibnamefont {Hessel}},\
  }\bibfield  {title} {\enquote {\bibinfo {title} {Micro-structured reactors
  for gas phase reactions},}\ }\href@noop {} {\bibfield  {journal} {\bibinfo
  {journal} {Chemical Engineering Journal}\ }\textbf {\bibinfo {volume} {98}},\
  \bibinfo {pages} {1--38} (\bibinfo {year} {2004})}\BibitemShut {NoStop}%
\bibitem [{\citenamefont {DeWitt}(1999)}]{DeWitt1999}%
  \BibitemOpen
  \bibfield  {author} {\bibinfo {author} {\bibfnamefont {S.~H.}\ \bibnamefont
  {DeWitt}},\ }\bibfield  {title} {\enquote {\bibinfo {title} {Micro reactors
  for chemical synthesis},}\ }\href@noop {} {\bibfield  {journal} {\bibinfo
  {journal} {Current Opinion in Chemical Biology}\ }\textbf {\bibinfo {volume}
  {3}},\ \bibinfo {pages} {350--356} (\bibinfo {year} {1999})}\BibitemShut
  {NoStop}%
\bibitem [{\citenamefont {Anastas}\ and\ \citenamefont
  {Eghbali}(2010)}]{Anastas2010}%
  \BibitemOpen
  \bibfield  {author} {\bibinfo {author} {\bibfnamefont {P.}~\bibnamefont
  {Anastas}}\ and\ \bibinfo {author} {\bibfnamefont {N.}~\bibnamefont
  {Eghbali}},\ }\bibfield  {title} {\enquote {\bibinfo {title} {Green
  chemistry: Principles and practice},}\ }\href {\doibase 10.1039/B918763B}
  {\bibfield  {journal} {\bibinfo  {journal} {Chem. Soc. Rev.}\ }\textbf
  {\bibinfo {volume} {39}},\ \bibinfo {pages} {301--312} (\bibinfo {year}
  {2010})}\BibitemShut {NoStop}%
\bibitem [{\citenamefont {Sengers}\ \emph {et~al.}(1971)\citenamefont
  {Sengers}, \citenamefont {Klein}, \citenamefont {Gallagher},\ and\
  \citenamefont {DIV.}}]{Sengers1971}%
  \BibitemOpen
  \bibfield  {author} {\bibinfo {author} {\bibfnamefont {J.}~\bibnamefont
  {Sengers}}, \bibinfo {author} {\bibfnamefont {M.}~\bibnamefont {Klein}},
  \bibinfo {author} {\bibfnamefont {J.}~\bibnamefont {Gallagher}}, \ and\
  \bibinfo {author} {\bibfnamefont {N.~B. O. S. W. D. C.~H.}\ \bibnamefont
  {DIV.}},\ }\href@noop {} {\emph {\bibinfo {title}
  {Pressure-Volume-Temperature Relationships of Gases Virial Coefficients}}}\
  (\bibinfo  {publisher} {Defense Technical Information Center},\ \bibinfo
  {year} {1971})\BibitemShut {NoStop}%
\bibitem [{\citenamefont {Lemmon}\ \emph {et~al.}()\citenamefont {Lemmon},
  \citenamefont {Bell}, \citenamefont {Huber},\ and\ \citenamefont
  {McLinden}}]{NIST}%
  \BibitemOpen
  \bibfield  {author} {\bibinfo {author} {\bibfnamefont {E.~W.}\ \bibnamefont
  {Lemmon}}, \bibinfo {author} {\bibfnamefont {I.~H.}\ \bibnamefont {Bell}},
  \bibinfo {author} {\bibfnamefont {M.~L.}\ \bibnamefont {Huber}}, \ and\
  \bibinfo {author} {\bibfnamefont {M.~O.}\ \bibnamefont {McLinden}},\
  }\href@noop {} {\enquote {\bibinfo {title} {Thermophysical properties of
  fluid systems},}\ }\bibinfo {howpublished} {NIST Chemistry WebBook, NIST
  Standard Reference Database Number 69, Eds. P.J. Linstrom and W.G. Mallard,
  National Institute of Standards and Technology, Gaithersburg MD, 2089},\
  \bibinfo {note} {(retrieved November 28, 2023)}\BibitemShut {NoStop}%
\bibitem [{\citenamefont {Le~Bellac}, \citenamefont {Mortessagne},\ and\
  \citenamefont {Batrouni}(2004)}]{LeBellac2004}%
  \BibitemOpen
  \bibfield  {author} {\bibinfo {author} {\bibfnamefont {M.}~\bibnamefont
  {Le~Bellac}}, \bibinfo {author} {\bibfnamefont {F.}~\bibnamefont
  {Mortessagne}}, \ and\ \bibinfo {author} {\bibfnamefont {G.~G.}\ \bibnamefont
  {Batrouni}},\ }\href@noop {} {\emph {\bibinfo {title} {Equilibrium and
  Non-Equilibrium Statistical Thermodynamics}}}\ (\bibinfo  {publisher}
  {Cambridge University Press},\ \bibinfo {year} {2004})\BibitemShut {NoStop}%
\bibitem [{\citenamefont {Sharma}\ and\ \citenamefont
  {Kumar}(2019)}]{Kumar2019}%
  \BibitemOpen
  \bibfield  {author} {\bibinfo {author} {\bibfnamefont {B.}~\bibnamefont
  {Sharma}}\ and\ \bibinfo {author} {\bibfnamefont {R.}~\bibnamefont {Kumar}},\
  }\bibfield  {title} {\enquote {\bibinfo {title} {Estimation of bulk viscosity
  of dilute gases using a nonequilibrium molecular dynamics approach},}\
  }\href@noop {} {\bibfield  {journal} {\bibinfo  {journal} {Phys. Rev. E}\
  }\textbf {\bibinfo {volume} {100}},\ \bibinfo {pages} {013309} (\bibinfo
  {year} {2019})}\BibitemShut {NoStop}%
\bibitem [{\citenamefont {Atkins}\ and\ \citenamefont
  {de~Paula}(2006)}]{book_Atkins}%
  \BibitemOpen
  \bibfield  {author} {\bibinfo {author} {\bibfnamefont {P.}~\bibnamefont
  {Atkins}}\ and\ \bibinfo {author} {\bibfnamefont {J.}~\bibnamefont
  {de~Paula}},\ }\href@noop {} {\emph {\bibinfo {title} {Atkins' Physical
  Chemistry}}}\ (\bibinfo  {publisher} {Oxford University Press},\ \bibinfo
  {year} {2006})\BibitemShut {NoStop}%
\bibitem [{Note1()}]{Note1}%
  \BibitemOpen
  \bibinfo {note} {It is worthwhile to note that $\delta (z)$ has unit
  meter$^{-1}$, as can be seen from how the integral $\DOTSI \intop \ilimits@
  \limits _{-a}^a \delta (z) dz =1$ is dimensionless (for any value of $a
  \protect \neq 0$) \cite {book_MathMeth}.}\BibitemShut {Stop}%
\bibitem [{\citenamefont {Lam}(2006)}]{Lam2006}%
  \BibitemOpen
  \bibfield  {author} {\bibinfo {author} {\bibfnamefont {S.~H.}\ \bibnamefont
  {Lam}},\ }\bibfield  {title} {\enquote {\bibinfo {title} {Multicomponent
  diffusion revisited},}\ }\href@noop {} {\bibfield  {journal} {\bibinfo
  {journal} {Physics of Fluids}\ }\textbf {\bibinfo {volume} {18}},\ \bibinfo
  {pages} {073101} (\bibinfo {year} {2006})}\BibitemShut {NoStop}%
\bibitem [{\citenamefont {Krishna}(1993)}]{Krishna1993}%
  \BibitemOpen
  \bibfield  {author} {\bibinfo {author} {\bibfnamefont {R.}~\bibnamefont
  {Krishna}},\ }\bibfield  {title} {\enquote {\bibinfo {title} {Problems and
  pitfalls in the use of the fick formulation for intraparticle diffusion},}\
  }\href@noop {} {\bibfield  {journal} {\bibinfo  {journal} {Chemical
  Engineering Science}\ }\textbf {\bibinfo {volume} {48}},\ \bibinfo {pages}
  {845--861} (\bibinfo {year} {1993})}\BibitemShut {NoStop}%
\bibitem [{\citenamefont {Kjelstrup}\ and\ \citenamefont
  {Bedeaux}(2020)}]{book_NEThermo}%
  \BibitemOpen
  \bibfield  {author} {\bibinfo {author} {\bibfnamefont {S.}~\bibnamefont
  {Kjelstrup}}\ and\ \bibinfo {author} {\bibfnamefont {D.}~\bibnamefont
  {Bedeaux}},\ }\href@noop {} {\emph {\bibinfo {title} {Non-equilibrium
  Thermodynamics of Heterogeneous Systems}}}\ (\bibinfo  {publisher} {WSPC},\
  \bibinfo {year} {2020})\BibitemShut {NoStop}%
\bibitem [{\citenamefont {Govender}\ and\ \citenamefont
  {Friedrich}(2017)}]{Govender2017}%
  \BibitemOpen
  \bibfield  {author} {\bibinfo {author} {\bibfnamefont {S.}~\bibnamefont
  {Govender}}\ and\ \bibinfo {author} {\bibfnamefont {H.~B.}\ \bibnamefont
  {Friedrich}},\ }\bibfield  {title} {\enquote {\bibinfo {title} {Monoliths: A
  review of the basics, preparation methods and their relevance to
  oxidation},}\ }\href@noop {} {\bibfield  {journal} {\bibinfo  {journal}
  {Catalysts}\ }\textbf {\bibinfo {volume} {7}} (\bibinfo {year}
  {2017})}\BibitemShut {NoStop}%
\bibitem [{\citenamefont {Moulijn}\ and\ \citenamefont
  {Kapteijn}(2013)}]{Kapteijn2013}%
  \BibitemOpen
  \bibfield  {author} {\bibinfo {author} {\bibfnamefont {J.~A.}\ \bibnamefont
  {Moulijn}}\ and\ \bibinfo {author} {\bibfnamefont {F.}~\bibnamefont
  {Kapteijn}},\ }\bibfield  {title} {\enquote {\bibinfo {title} {Monolithic
  reactors in catalysis: excellent control},}\ }\href@noop {} {\bibfield
  {journal} {\bibinfo  {journal} {Curr. Opin. Chem. Eng.}\ }\textbf {\bibinfo
  {volume} {2}},\ \bibinfo {pages} {346--353} (\bibinfo {year} {2013})},\
  \bibinfo {note} {energy and environmental engineering / Reaction engineering
  and catalysis}\BibitemShut {NoStop}%
\bibitem [{\citenamefont {Zwanzig}(1992)}]{Zwanzig1992}%
  \BibitemOpen
  \bibfield  {author} {\bibinfo {author} {\bibfnamefont {R.}~\bibnamefont
  {Zwanzig}},\ }\bibfield  {title} {\enquote {\bibinfo {title} {Diffusion past
  an entropy barrier},}\ }\href@noop {} {\bibfield  {journal} {\bibinfo
  {journal} {J. Phys. Chem.}\ }\textbf {\bibinfo {volume} {96}},\ \bibinfo
  {pages} {3926} (\bibinfo {year} {1992})}\BibitemShut {NoStop}%
\bibitem [{\citenamefont {Reguera}\ and\ \citenamefont
  {Rubi}(2001)}]{Reguera2001}%
  \BibitemOpen
  \bibfield  {author} {\bibinfo {author} {\bibfnamefont {D.}~\bibnamefont
  {Reguera}}\ and\ \bibinfo {author} {\bibfnamefont {J.~M.}\ \bibnamefont
  {Rubi}},\ }\bibfield  {title} {\enquote {\bibinfo {title} {Kinetic equations
  for diffusion in the presence of entropic barriers},}\ }\href@noop {}
  {\bibfield  {journal} {\bibinfo  {journal} {Phys. Rev. E}\ }\textbf {\bibinfo
  {volume} {64}},\ \bibinfo {pages} {061106} (\bibinfo {year}
  {2001})}\BibitemShut {NoStop}%
\bibitem [{\citenamefont {Kalinay}\ and\ \citenamefont
  {Percus}(2005)}]{Kalinay2005}%
  \BibitemOpen
  \bibfield  {author} {\bibinfo {author} {\bibfnamefont {P.}~\bibnamefont
  {Kalinay}}\ and\ \bibinfo {author} {\bibfnamefont {J.~K.~P.}\ \bibnamefont
  {Percus}},\ }\bibfield  {title} {\enquote {\bibinfo {title} {Projection of
  two-dimensional diffusion in a narrow channel onto the longitudinal
  dimension},}\ }\href {\doibase https://doi.org/10.1063/1.1899150} {\bibfield
  {journal} {\bibinfo  {journal} {J. Chem. Phys.}\ }\textbf {\bibinfo {volume}
  {122}},\ \bibinfo {pages} {204701} (\bibinfo {year} {2005})}\BibitemShut
  {NoStop}%
\bibitem [{\citenamefont {Kalinay}\ and\ \citenamefont
  {Percus}(2008)}]{Kalinay2008}%
  \BibitemOpen
  \bibfield  {author} {\bibinfo {author} {\bibfnamefont {P.}~\bibnamefont
  {Kalinay}}\ and\ \bibinfo {author} {\bibfnamefont {J.~K.}\ \bibnamefont
  {Percus}},\ }\bibfield  {title} {\enquote {\bibinfo {title} {Approximations
  of the generalized fick-jacobs equation},}\ }\href {\doibase
  10.1103/PhysRevE.78.021103} {\bibfield  {journal} {\bibinfo  {journal} {Phys.
  Rev. E}\ }\textbf {\bibinfo {volume} {78}},\ \bibinfo {pages} {021103}
  (\bibinfo {year} {2008})}\BibitemShut {NoStop}%
\bibitem [{\citenamefont {Martens}\ \emph {et~al.}(2011)\citenamefont
  {Martens}, \citenamefont {Schmid}, \citenamefont {Schimansky-Geier},\ and\
  \citenamefont {H\"anggi}}]{Martens2011}%
  \BibitemOpen
  \bibfield  {author} {\bibinfo {author} {\bibfnamefont {S.}~\bibnamefont
  {Martens}}, \bibinfo {author} {\bibfnamefont {G.}~\bibnamefont {Schmid}},
  \bibinfo {author} {\bibfnamefont {L.}~\bibnamefont {Schimansky-Geier}}, \
  and\ \bibinfo {author} {\bibfnamefont {P.}~\bibnamefont {H\"anggi}},\
  }\bibfield  {title} {\enquote {\bibinfo {title} {Entropic particle transport:
  Higher-order corrections to the {F}ick-{J}acobs diffusion equation},}\ }\href
  {\doibase 10.1103/PhysRevE.83.051135} {\bibfield  {journal} {\bibinfo
  {journal} {Phys. Rev. E}\ }\textbf {\bibinfo {volume} {83}},\ \bibinfo
  {pages} {051135} (\bibinfo {year} {2011})}\BibitemShut {NoStop}%
\bibitem [{\citenamefont {Chac\'on-Acosta}, \citenamefont {Pineda},\ and\
  \citenamefont {Dagdug}(2013)}]{Dagdug2013}%
  \BibitemOpen
  \bibfield  {author} {\bibinfo {author} {\bibfnamefont {G.}~\bibnamefont
  {Chac\'on-Acosta}}, \bibinfo {author} {\bibfnamefont {I.}~\bibnamefont
  {Pineda}}, \ and\ \bibinfo {author} {\bibfnamefont {L.}~\bibnamefont
  {Dagdug}},\ }\bibfield  {title} {\enquote {\bibinfo {title} {Diffusion in
  narrow channels on curved manifolds},}\ }\href {\doibase
  http://dx.doi.org/10.1063/1.4836617} {\bibfield  {journal} {\bibinfo
  {journal} {J. Chem. Phys.}\ }\textbf {\bibinfo {volume} {139}},\ \bibinfo
  {eid} {214115} (\bibinfo {year} {2013})}\BibitemShut {NoStop}%
\bibitem [{\citenamefont {Malgaretti}\ and\ \citenamefont
  {Harting}(2023)}]{Malgaretti2023}%
  \BibitemOpen
  \bibfield  {author} {\bibinfo {author} {\bibfnamefont {P.}~\bibnamefont
  {Malgaretti}}\ and\ \bibinfo {author} {\bibfnamefont {J.}~\bibnamefont
  {Harting}},\ }\bibfield  {title} {\enquote {\bibinfo {title} {Closed formula
  for transport across constrictions},}\ }\href {\doibase 10.3390/e25030470}
  {\bibfield  {journal} {\bibinfo  {journal} {Entropy}\ }\textbf {\bibinfo
  {volume} {25}},\ \bibinfo {pages} {470} (\bibinfo {year} {2023})}\BibitemShut
  {NoStop}%
\bibitem [{\citenamefont {Reguera}\ \emph {et~al.}(2006)\citenamefont
  {Reguera}, \citenamefont {Schmid}, \citenamefont {Burada}, \citenamefont
  {Rubi}, \citenamefont {Reimann},\ and\ \citenamefont
  {H\"anggi}}]{Reguera2006}%
  \BibitemOpen
  \bibfield  {author} {\bibinfo {author} {\bibfnamefont {D.}~\bibnamefont
  {Reguera}}, \bibinfo {author} {\bibfnamefont {G.}~\bibnamefont {Schmid}},
  \bibinfo {author} {\bibfnamefont {P.~S.}\ \bibnamefont {Burada}}, \bibinfo
  {author} {\bibfnamefont {J.~M.}\ \bibnamefont {Rubi}}, \bibinfo {author}
  {\bibfnamefont {P.}~\bibnamefont {Reimann}}, \ and\ \bibinfo {author}
  {\bibfnamefont {P.}~\bibnamefont {H\"anggi}},\ }\bibfield  {title} {\enquote
  {\bibinfo {title} {Entropic transport: Kinetics, scaling, and control
  mechanisms},}\ }\href {\doibase 10.1103/PhysRevLett.96.130603} {\bibfield
  {journal} {\bibinfo  {journal} {Phys. Rev. Lett.}\ }\textbf {\bibinfo
  {volume} {96}},\ \bibinfo {pages} {130603} (\bibinfo {year}
  {2006})}\BibitemShut {NoStop}%
\bibitem [{\citenamefont {Reguera}\ \emph {et~al.}(2012)\citenamefont
  {Reguera}, \citenamefont {Luque}, \citenamefont {Burada}, \citenamefont
  {Schmid}, \citenamefont {Rubi},\ and\ \citenamefont
  {H\"anggi}}]{Reguera2012}%
  \BibitemOpen
  \bibfield  {author} {\bibinfo {author} {\bibfnamefont {D.}~\bibnamefont
  {Reguera}}, \bibinfo {author} {\bibfnamefont {A.}~\bibnamefont {Luque}},
  \bibinfo {author} {\bibfnamefont {P.~S.}\ \bibnamefont {Burada}}, \bibinfo
  {author} {\bibfnamefont {G.}~\bibnamefont {Schmid}}, \bibinfo {author}
  {\bibfnamefont {J.~M.}\ \bibnamefont {Rubi}}, \ and\ \bibinfo {author}
  {\bibfnamefont {P.}~\bibnamefont {H\"anggi}},\ }\bibfield  {title} {\enquote
  {\bibinfo {title} {Entropic splitter for particle separation},}\ }\href
  {\doibase 10.1103/PhysRevLett.108.020604} {\bibfield  {journal} {\bibinfo
  {journal} {Phys. Rev. Lett.}\ }\textbf {\bibinfo {volume} {108}},\ \bibinfo
  {pages} {020604} (\bibinfo {year} {2012})}\BibitemShut {NoStop}%
\bibitem [{\citenamefont {Marini Bettolo~Marconi}, \citenamefont {Malgaretti},\
  and\ \citenamefont {Pagonabarraga}(2015)}]{Marconi2015}%
  \BibitemOpen
  \bibfield  {author} {\bibinfo {author} {\bibfnamefont {U.}~\bibnamefont
  {Marini Bettolo~Marconi}}, \bibinfo {author} {\bibfnamefont {P.}~\bibnamefont
  {Malgaretti}}, \ and\ \bibinfo {author} {\bibfnamefont {I.}~\bibnamefont
  {Pagonabarraga}},\ }\bibfield  {title} {\enquote {\bibinfo {title} {Tracer
  diffusion of hard-sphere binary mixtures under nano-confinement},}\ }\href
  {\doibase 10.1063/1.4934994} {\bibfield  {journal} {\bibinfo  {journal} {J.
  Chem. Phys.}\ }\textbf {\bibinfo {volume} {143}},\ \bibinfo {pages} {184501}
  (\bibinfo {year} {2015})}\BibitemShut {NoStop}%
\bibitem [{\citenamefont {Malgaretti}, \citenamefont {Pagonabarraga},\ and\
  \citenamefont {Rubi}(2016)}]{Malgaretti2016_entropy}%
  \BibitemOpen
  \bibfield  {author} {\bibinfo {author} {\bibfnamefont {P.}~\bibnamefont
  {Malgaretti}}, \bibinfo {author} {\bibfnamefont {I.}~\bibnamefont
  {Pagonabarraga}}, \ and\ \bibinfo {author} {\bibfnamefont {J.}~\bibnamefont
  {Rubi}},\ }\bibfield  {title} {\enquote {\bibinfo {title} {Rectification and
  non-gaussian diffusion in heterogeneous media},}\ }\href@noop {} {\bibfield
  {journal} {\bibinfo  {journal} {Entropy}\ }\textbf {\bibinfo {volume} {18}},\
  \bibinfo {pages} {394} (\bibinfo {year} {2016})}\BibitemShut {NoStop}%
\bibitem [{\citenamefont {Puertas}, \citenamefont {Malgaretti},\ and\
  \citenamefont {Pagonabarraga}(2018)}]{Puertas2018}%
  \BibitemOpen
  \bibfield  {author} {\bibinfo {author} {\bibfnamefont {A.}~\bibnamefont
  {Puertas}}, \bibinfo {author} {\bibfnamefont {P.}~\bibnamefont {Malgaretti}},
  \ and\ \bibinfo {author} {\bibfnamefont {I.}~\bibnamefont {Pagonabarraga}},\
  }\bibfield  {title} {\enquote {\bibinfo {title} {Active microrheology in
  corrugated channels},}\ }\href@noop {} {\bibfield  {journal} {\bibinfo
  {journal} {J. Chem. Phys.}\ }\textbf {\bibinfo {volume} {149}},\ \bibinfo
  {pages} {174908} (\bibinfo {year} {2018})}\BibitemShut {NoStop}%
\bibitem [{\citenamefont {Bianco}\ and\ \citenamefont
  {Malgaretti}(2016)}]{Bianco2016}%
  \BibitemOpen
  \bibfield  {author} {\bibinfo {author} {\bibfnamefont {V.}~\bibnamefont
  {Bianco}}\ and\ \bibinfo {author} {\bibfnamefont {P.}~\bibnamefont
  {Malgaretti}},\ }\bibfield  {title} {\enquote {\bibinfo {title}
  {Non-monotonous polymer translocation time across corrugated channels:
  Comparison between fick-jacobs approximation and numerical simulations},}\
  }\href {\doibase 10.1063/1.4961697} {\bibfield  {journal} {\bibinfo
  {journal} {J. Chem. Phys.}\ }\textbf {\bibinfo {volume} {145}},\ \bibinfo
  {pages} {114904} (\bibinfo {year} {2016})}\BibitemShut {NoStop}%
\bibitem [{\citenamefont {Locatelli}\ \emph {et~al.}(2023)\citenamefont
  {Locatelli}, \citenamefont {Bianco}, \citenamefont {Valeriani},\ and\
  \citenamefont {Malgaretti}}]{Locatelli2023}%
  \BibitemOpen
  \bibfield  {author} {\bibinfo {author} {\bibfnamefont {E.}~\bibnamefont
  {Locatelli}}, \bibinfo {author} {\bibfnamefont {V.}~\bibnamefont {Bianco}},
  \bibinfo {author} {\bibfnamefont {C.}~\bibnamefont {Valeriani}}, \ and\
  \bibinfo {author} {\bibfnamefont {P.}~\bibnamefont {Malgaretti}},\
  }\href@noop {} {\bibfield  {journal} {\bibinfo  {journal} {Phys. Rev. Lett.}\
  }\textbf {\bibinfo {volume} {131}},\ \bibinfo {pages} {048101} (\bibinfo
  {year} {2023})}\BibitemShut {NoStop}%
\bibitem [{\citenamefont {Malgaretti}, \citenamefont {Pagonabarraga},\ and\
  \citenamefont {Rubi}(2014)}]{Malgaretti2014}%
  \BibitemOpen
  \bibfield  {author} {\bibinfo {author} {\bibfnamefont {P.}~\bibnamefont
  {Malgaretti}}, \bibinfo {author} {\bibfnamefont {I.}~\bibnamefont
  {Pagonabarraga}}, \ and\ \bibinfo {author} {\bibfnamefont {J.~M.}\
  \bibnamefont {Rubi}},\ }\bibfield  {title} {\enquote {\bibinfo {title}
  {Entropic electrokinetics},}\ }\href@noop {} {\bibfield  {journal} {\bibinfo
  {journal} {Phys. Rev. Lett}\ }\textbf {\bibinfo {volume} {113}},\ \bibinfo
  {pages} {128301} (\bibinfo {year} {2014})}\BibitemShut {NoStop}%
\bibitem [{\citenamefont {Chinappi}\ and\ \citenamefont
  {Malgaretti}(2018)}]{Chinappi2018}%
  \BibitemOpen
  \bibfield  {author} {\bibinfo {author} {\bibfnamefont {M.}~\bibnamefont
  {Chinappi}}\ and\ \bibinfo {author} {\bibfnamefont {P.}~\bibnamefont
  {Malgaretti}},\ }\bibfield  {title} {\enquote {\bibinfo {title} {Charge
  polarization, local electroneutrality breakdown and eddy formation due to
  electroosmosis in varying-section channels},}\ }\href {\doibase
  10.1039/C8SM01298A} {\bibfield  {journal} {\bibinfo  {journal} {Soft Matter}\
  }\textbf {\bibinfo {volume} {14}},\ \bibinfo {pages} {9083} (\bibinfo {year}
  {2018})}\BibitemShut {NoStop}%
\bibitem [{\citenamefont {Malgaretti}\ \emph {et~al.}(2019)\citenamefont
  {Malgaretti}, \citenamefont {Janssen}, \citenamefont {Pagonabarraga},\ and\
  \citenamefont {Rubi}}]{Malgaretti2019_JCP}%
  \BibitemOpen
  \bibfield  {author} {\bibinfo {author} {\bibfnamefont {P.}~\bibnamefont
  {Malgaretti}}, \bibinfo {author} {\bibfnamefont {M.}~\bibnamefont {Janssen}},
  \bibinfo {author} {\bibfnamefont {I.}~\bibnamefont {Pagonabarraga}}, \ and\
  \bibinfo {author} {\bibfnamefont {J.~M.}\ \bibnamefont {Rubi}},\ }\bibfield
  {title} {\enquote {\bibinfo {title} {Driving an electrolyte through a
  corrugated nanopore},}\ }\href {\doibase 10.1063/1.5110349} {\bibfield
  {journal} {\bibinfo  {journal} {J. Chem. Phys.}\ }\textbf {\bibinfo {volume}
  {151}},\ \bibinfo {pages} {084902} (\bibinfo {year} {2019})}\BibitemShut
  {NoStop}%
\bibitem [{\citenamefont {Sandoval}\ and\ \citenamefont
  {Dagdug}(2014)}]{Dagdug2014}%
  \BibitemOpen
  \bibfield  {author} {\bibinfo {author} {\bibfnamefont {M.}~\bibnamefont
  {Sandoval}}\ and\ \bibinfo {author} {\bibfnamefont {L.}~\bibnamefont
  {Dagdug}},\ }\bibfield  {title} {\enquote {\bibinfo {title} {Effective
  diffusion of confined active brownian swimmers},}\ }\href {\doibase
  10.1103/PhysRevE.90.062711} {\bibfield  {journal} {\bibinfo  {journal} {Phys.
  Rev. E}\ }\textbf {\bibinfo {volume} {90}},\ \bibinfo {pages} {062711}
  (\bibinfo {year} {2014})}\BibitemShut {NoStop}%
\bibitem [{\citenamefont {Kalinay}(2022)}]{Kalinay2022}%
  \BibitemOpen
  \bibfield  {author} {\bibinfo {author} {\bibfnamefont {P.}~\bibnamefont
  {Kalinay}},\ }\bibfield  {title} {\enquote {\bibinfo {title} {Transverse
  dichotomic ratchet in a two-dimensional corrugated channel},}\ }\href
  {\doibase 10.1103/PhysRevE.106.044126} {\bibfield  {journal} {\bibinfo
  {journal} {Phys. Rev. E}\ }\textbf {\bibinfo {volume} {106}},\ \bibinfo
  {pages} {044126} (\bibinfo {year} {2022})}\BibitemShut {NoStop}%
\bibitem [{\citenamefont {Antunes}\ \emph {et~al.}(2022)\citenamefont
  {Antunes}, \citenamefont {Malgaretti}, \citenamefont {Harting},\ and\
  \citenamefont {Dietrich}}]{Antunes2022}%
  \BibitemOpen
  \bibfield  {author} {\bibinfo {author} {\bibfnamefont {G.~C.}\ \bibnamefont
  {Antunes}}, \bibinfo {author} {\bibfnamefont {P.}~\bibnamefont {Malgaretti}},
  \bibinfo {author} {\bibfnamefont {J.}~\bibnamefont {Harting}}, \ and\
  \bibinfo {author} {\bibfnamefont {S.}~\bibnamefont {Dietrich}},\ }\bibfield
  {title} {\enquote {\bibinfo {title} {Pumping and mixing in active pores},}\
  }\href@noop {} {\bibfield  {journal} {\bibinfo  {journal} {Phys. Rev. Lett.}\
  }\textbf {\bibinfo {volume} {129}},\ \bibinfo {pages} {188003} (\bibinfo
  {year} {2022})}\BibitemShut {NoStop}%
\bibitem [{\citenamefont {Antunes}, \citenamefont {Malgaretti},\ and\
  \citenamefont {Harting}(2023)}]{Antunes2023}%
  \BibitemOpen
  \bibfield  {author} {\bibinfo {author} {\bibfnamefont {G.~C.}\ \bibnamefont
  {Antunes}}, \bibinfo {author} {\bibfnamefont {P.}~\bibnamefont {Malgaretti}},
  \ and\ \bibinfo {author} {\bibfnamefont {J.}~\bibnamefont {Harting}},\
  }\bibfield  {title} {\enquote {\bibinfo {title} {{Turning catalytically
  active pores into active pumps}},}\ }\href {\doibase 10.1063/5.0160414}
  {\bibfield  {journal} {\bibinfo  {journal} {The Journal of Chemical Physics}\
  }\textbf {\bibinfo {volume} {159}},\ \bibinfo {pages} {134903} (\bibinfo
  {year} {2023})},\ \Eprint
  {http://arxiv.org/abs/https://pubs.aip.org/aip/jcp/article-pdf/doi/10.1063/5.0160414/18149640/134903\_1\_5.0160414.pdf}
  {https://pubs.aip.org/aip/jcp/article-pdf/doi/10.1063/5.0160414/18149640/134903\_1\_5.0160414.pdf}
  \BibitemShut {NoStop}%
\bibitem [{\citenamefont {Schlichting}(1979)}]{Schlichting1979}%
  \BibitemOpen
  \bibfield  {author} {\bibinfo {author} {\bibfnamefont {H.}~\bibnamefont
  {Schlichting}},\ }\href@noop {} {\emph {\bibinfo {title} {Boundary Layer
  Theory}}}\ (\bibinfo  {publisher} {McGraw-Hill},\ \bibinfo {year}
  {1979})\BibitemShut {NoStop}%
\bibitem [{\citenamefont {Cramer}(2012)}]{Cramer2012}%
  \BibitemOpen
  \bibfield  {author} {\bibinfo {author} {\bibfnamefont {M.~S.}\ \bibnamefont
  {Cramer}},\ }\bibfield  {title} {\enquote {\bibinfo {title} {{Numerical
  estimates for the bulk viscosity of ideal gases}},}\ }\href@noop {}
  {\bibfield  {journal} {\bibinfo  {journal} {Physics of Fluids}\ }\textbf
  {\bibinfo {volume} {24}},\ \bibinfo {pages} {066102} (\bibinfo {year}
  {2012})}\BibitemShut {NoStop}%
\bibitem [{\citenamefont {Brekhovskikh}\ and\ \citenamefont
  {Goncharov}(1994)}]{book_Russianguy}%
  \BibitemOpen
  \bibfield  {author} {\bibinfo {author} {\bibfnamefont {L.~M.}\ \bibnamefont
  {Brekhovskikh}}\ and\ \bibinfo {author} {\bibfnamefont {V.}~\bibnamefont
  {Goncharov}},\ }\href@noop {} {\emph {\bibinfo {title} {Mechanics of Continua
  and Wave Dynamics}}},\ Springer Series on Wave Phenomena\ (\bibinfo
  {publisher} {Springer},\ \bibinfo {year} {1994})\BibitemShut {NoStop}%
\bibitem [{\citenamefont {Kadoya}, \citenamefont {Matsunaga},\ and\
  \citenamefont {Nagashima}(1985)}]{Nagashima1985}%
  \BibitemOpen
  \bibfield  {author} {\bibinfo {author} {\bibfnamefont {K.}~\bibnamefont
  {Kadoya}}, \bibinfo {author} {\bibfnamefont {N.}~\bibnamefont {Matsunaga}}, \
  and\ \bibinfo {author} {\bibfnamefont {A.}~\bibnamefont {Nagashima}},\
  }\bibfield  {title} {\enquote {\bibinfo {title} {Viscosity and thermal
  conductivity of dry air in the gaseous phase},}\ }\href@noop {} {\bibfield
  {journal} {\bibinfo  {journal} {Journal of Physical and Chemical Reference
  Data}\ }\textbf {\bibinfo {volume} {14}},\ \bibinfo {pages} {947--970}
  (\bibinfo {year} {1985})}\BibitemShut {NoStop}%
\bibitem [{\citenamefont {Wilke}\ and\ \citenamefont {Lee}(1955)}]{Wilke1955}%
  \BibitemOpen
  \bibfield  {author} {\bibinfo {author} {\bibfnamefont {C.~R.}\ \bibnamefont
  {Wilke}}\ and\ \bibinfo {author} {\bibfnamefont {C.~Y.}\ \bibnamefont
  {Lee}},\ }\bibfield  {title} {\enquote {\bibinfo {title} {Estimation of
  diffusion coefficients for gases and vapors},}\ }\href {\doibase
  10.1021/ie50546a056} {\bibfield  {journal} {\bibinfo  {journal} {Industrial
  {\&} Engineering Chemistry}\ }\textbf {\bibinfo {volume} {47}},\ \bibinfo
  {pages} {1253--1257} (\bibinfo {year} {1955})}\BibitemShut {NoStop}%
\bibitem [{\citenamefont {Reinke}\ \emph {et~al.}(2004)\citenamefont {Reinke},
  \citenamefont {Mantzaras}, \citenamefont {Schaeren}, \citenamefont {Bombach},
  \citenamefont {Inauen},\ and\ \citenamefont {Schenker}}]{Reinke2004}%
  \BibitemOpen
  \bibfield  {author} {\bibinfo {author} {\bibfnamefont {M.}~\bibnamefont
  {Reinke}}, \bibinfo {author} {\bibfnamefont {J.}~\bibnamefont {Mantzaras}},
  \bibinfo {author} {\bibfnamefont {R.}~\bibnamefont {Schaeren}}, \bibinfo
  {author} {\bibfnamefont {R.}~\bibnamefont {Bombach}}, \bibinfo {author}
  {\bibfnamefont {A.}~\bibnamefont {Inauen}}, \ and\ \bibinfo {author}
  {\bibfnamefont {S.}~\bibnamefont {Schenker}},\ }\bibfield  {title} {\enquote
  {\bibinfo {title} {High-pressure catalytic combustion of methane over
  platinum: In situ experiments and detailed numerical predictions},}\
  }\href@noop {} {\bibfield  {journal} {\bibinfo  {journal} {Combustion and
  Flame}\ }\textbf {\bibinfo {volume} {136}},\ \bibinfo {pages} {217--240}
  (\bibinfo {year} {2004})}\BibitemShut {NoStop}%
\bibitem [{\citenamefont {Papp}\ and\ \citenamefont
  {Steinrück}(2013)}]{Papp2013}%
  \BibitemOpen
  \bibfield  {author} {\bibinfo {author} {\bibfnamefont {C.}~\bibnamefont
  {Papp}}\ and\ \bibinfo {author} {\bibfnamefont {H.-P.}\ \bibnamefont
  {Steinrück}},\ }\bibfield  {title} {\enquote {\bibinfo {title} {In situ
  high-resolution x-ray photoelectron spectroscopy – fundamental insights in
  surface reactions},}\ }\href@noop {} {\bibfield  {journal} {\bibinfo
  {journal} {Surface Science Reports}\ }\textbf {\bibinfo {volume} {68}},\
  \bibinfo {pages} {446--487} (\bibinfo {year} {2013})}\BibitemShut {NoStop}%
\bibitem [{\citenamefont {Ghavam}\ \emph {et~al.}(2021)\citenamefont {Ghavam},
  \citenamefont {Vahdati}, \citenamefont {Wilson},\ and\ \citenamefont
  {Styring}}]{Ghavam2021}%
  \BibitemOpen
  \bibfield  {author} {\bibinfo {author} {\bibfnamefont {S.}~\bibnamefont
  {Ghavam}}, \bibinfo {author} {\bibfnamefont {M.}~\bibnamefont {Vahdati}},
  \bibinfo {author} {\bibfnamefont {I.~A.~G.}\ \bibnamefont {Wilson}}, \ and\
  \bibinfo {author} {\bibfnamefont {P.}~\bibnamefont {Styring}},\ }\bibfield
  {title} {\enquote {\bibinfo {title} {Sustainable ammonia production
  processes},}\ }\href {\doibase 10.3389/fenrg.2021.580808} {\bibfield
  {journal} {\bibinfo  {journal} {Frontiers in Energy Research}\ }\textbf
  {\bibinfo {volume} {9}} (\bibinfo {year} {2021}),\
  10.3389/fenrg.2021.580808}\BibitemShut {NoStop}%
\bibitem [{\citenamefont {Xu}, \citenamefont {Liu},\ and\ \citenamefont
  {Wang}(2009)}]{Xu2009}%
  \BibitemOpen
  \bibfield  {author} {\bibinfo {author} {\bibfnamefont {G.}~\bibnamefont
  {Xu}}, \bibinfo {author} {\bibfnamefont {R.}~\bibnamefont {Liu}}, \ and\
  \bibinfo {author} {\bibfnamefont {J.}~\bibnamefont {Wang}},\ }\bibfield
  {title} {\enquote {\bibinfo {title} {Electrochemical synthesis of ammonia
  using a cell with a nafion membrane and smfe0.7cu0.3-xnixo3 (x = 0-0.3)
  cathode at atmospheric pressure and lower temperature},}\ }\href@noop {}
  {\bibfield  {journal} {\bibinfo  {journal} {Science in China Series B:
  Chemistry}\ }\textbf {\bibinfo {volume} {52}},\ \bibinfo {pages} {1171--1175}
  (\bibinfo {year} {2009})}\BibitemShut {NoStop}%
\bibitem [{\citenamefont {Colin}, \citenamefont {Lalonde},\ and\ \citenamefont
  {Caen}(2004)}]{Colin2004}%
  \BibitemOpen
  \bibfield  {author} {\bibinfo {author} {\bibfnamefont {S.}~\bibnamefont
  {Colin}}, \bibinfo {author} {\bibfnamefont {P.}~\bibnamefont {Lalonde}}, \
  and\ \bibinfo {author} {\bibfnamefont {R.}~\bibnamefont {Caen}},\ }\bibfield
  {title} {\enquote {\bibinfo {title} {Validation of a second-order slip flow
  model in rectangular microchannels},}\ }\href {\doibase
  10.1080/01457630490280047} {\bibfield  {journal} {\bibinfo  {journal} {Heat
  Transfer Engineering}\ }\textbf {\bibinfo {volume} {25}},\ \bibinfo {pages}
  {23 – 30} (\bibinfo {year} {2004})},\ \bibinfo {note} {cited by: 249; All
  Open Access, Bronze Open Access, Green Open Access}\BibitemShut {NoStop}%
\bibitem [{\citenamefont {Jennings}(1988)}]{Jennings1988}%
  \BibitemOpen
  \bibfield  {author} {\bibinfo {author} {\bibfnamefont {S.}~\bibnamefont
  {Jennings}},\ }\bibfield  {title} {\enquote {\bibinfo {title} {The mean free
  path in air},}\ }\href {\doibase
  https://doi.org/10.1016/0021-8502(88)90219-4} {\bibfield  {journal} {\bibinfo
   {journal} {Journal of Aerosol Science}\ }\textbf {\bibinfo {volume} {19}},\
  \bibinfo {pages} {159--166} (\bibinfo {year} {1988})}\BibitemShut {NoStop}%
\bibitem [{\citenamefont {Balcombe}\ \emph {et~al.}(2017)\citenamefont
  {Balcombe}, \citenamefont {Anderson}, \citenamefont {Speirs}, \citenamefont
  {Brandon},\ and\ \citenamefont {Hawkes}}]{Balcombe2017}%
  \BibitemOpen
  \bibfield  {author} {\bibinfo {author} {\bibfnamefont {P.}~\bibnamefont
  {Balcombe}}, \bibinfo {author} {\bibfnamefont {K.}~\bibnamefont {Anderson}},
  \bibinfo {author} {\bibfnamefont {J.}~\bibnamefont {Speirs}}, \bibinfo
  {author} {\bibfnamefont {N.}~\bibnamefont {Brandon}}, \ and\ \bibinfo
  {author} {\bibfnamefont {A.}~\bibnamefont {Hawkes}},\ }\bibfield  {title}
  {\enquote {\bibinfo {title} {The natural gas supply chain: The importance of
  methane and carbon dioxide emissions},}\ }\href {\doibase
  10.1021/acssuschemeng.6b00144} {\bibfield  {journal} {\bibinfo  {journal}
  {ACS Sustainable Chemistry {\&} Engineering}\ }\textbf {\bibinfo {volume}
  {5}},\ \bibinfo {pages} {3--20} (\bibinfo {year} {2017})}\BibitemShut
  {NoStop}%
\bibitem [{\citenamefont {Raub}\ \emph {et~al.}(2000)\citenamefont {Raub},
  \citenamefont {Mathieu-Nolf}, \citenamefont {Hampson},\ and\ \citenamefont
  {Thom}}]{Raub2000}%
  \BibitemOpen
  \bibfield  {author} {\bibinfo {author} {\bibfnamefont {J.~A.}\ \bibnamefont
  {Raub}}, \bibinfo {author} {\bibfnamefont {M.}~\bibnamefont {Mathieu-Nolf}},
  \bibinfo {author} {\bibfnamefont {N.~B.}\ \bibnamefont {Hampson}}, \ and\
  \bibinfo {author} {\bibfnamefont {S.~R.}\ \bibnamefont {Thom}},\ }\bibfield
  {title} {\enquote {\bibinfo {title} {Carbon monoxide poisoning — a public
  health perspective},}\ }\href {\doibase
  https://doi.org/10.1016/S0300-483X(99)00217-6} {\bibfield  {journal}
  {\bibinfo  {journal} {Toxicology}\ }\textbf {\bibinfo {volume} {145}},\
  \bibinfo {pages} {1--14} (\bibinfo {year} {2000})}\BibitemShut {NoStop}%
\bibitem [{\citenamefont {Zoccali}, \citenamefont {Tranchida},\ and\
  \citenamefont {Mondello}(2019)}]{Zoccali2019}%
  \BibitemOpen
  \bibfield  {author} {\bibinfo {author} {\bibfnamefont {M.}~\bibnamefont
  {Zoccali}}, \bibinfo {author} {\bibfnamefont {P.~Q.}\ \bibnamefont
  {Tranchida}}, \ and\ \bibinfo {author} {\bibfnamefont {L.}~\bibnamefont
  {Mondello}},\ }\bibfield  {title} {\enquote {\bibinfo {title} {Fast gas
  chromatography-mass spectrometry: A review of the last decade},}\ }\href
  {\doibase https://doi.org/10.1016/j.trac.2019.06.006} {\bibfield  {journal}
  {\bibinfo  {journal} {TrAC Trends in Analytical Chemistry}\ }\textbf
  {\bibinfo {volume} {118}},\ \bibinfo {pages} {444--452} (\bibinfo {year}
  {2019})}\BibitemShut {NoStop}%
\bibitem [{\citenamefont {de~Zeeuw}\ \emph {et~al.}(2000)\citenamefont
  {de~Zeeuw}, \citenamefont {Peene}, \citenamefont {Jansen},\ and\
  \citenamefont {Lou}}]{Jaap2000}%
  \BibitemOpen
  \bibfield  {author} {\bibinfo {author} {\bibfnamefont {J.}~\bibnamefont
  {de~Zeeuw}}, \bibinfo {author} {\bibfnamefont {J.}~\bibnamefont {Peene}},
  \bibinfo {author} {\bibfnamefont {H.-G.}\ \bibnamefont {Jansen}}, \ and\
  \bibinfo {author} {\bibfnamefont {X.}~\bibnamefont {Lou}},\ }\bibfield
  {title} {\enquote {\bibinfo {title} {A simple way to speed up separations by
  gc-ms using short 0.53 mm columns and vacuum outlet conditions},}\
  }\href@noop {} {\bibfield  {journal} {\bibinfo  {journal} {Journal of High
  Resolution Chromatography}\ }\textbf {\bibinfo {volume} {23}},\ \bibinfo
  {pages} {677--680} (\bibinfo {year} {2000})}\BibitemShut {NoStop}%
\bibitem [{\citenamefont {Englhard}\ \emph {et~al.}(2021)\citenamefont
  {Englhard}, \citenamefont {Cao}, \citenamefont {Bochmann}, \citenamefont
  {Barr}, \citenamefont {Cadot}, \citenamefont {Quadrelli},\ and\ \citenamefont
  {Bachmann}}]{Englhard2021}%
  \BibitemOpen
  \bibfield  {author} {\bibinfo {author} {\bibfnamefont {J.}~\bibnamefont
  {Englhard}}, \bibinfo {author} {\bibfnamefont {Y.}~\bibnamefont {Cao}},
  \bibinfo {author} {\bibfnamefont {S.}~\bibnamefont {Bochmann}}, \bibinfo
  {author} {\bibfnamefont {M.~K.~S.}\ \bibnamefont {Barr}}, \bibinfo {author}
  {\bibfnamefont {S.}~\bibnamefont {Cadot}}, \bibinfo {author} {\bibfnamefont
  {E.~A.}\ \bibnamefont {Quadrelli}}, \ and\ \bibinfo {author} {\bibfnamefont
  {J.}~\bibnamefont {Bachmann}},\ }\bibfield  {title} {\enquote {\bibinfo
  {title} {Stabilizing an ultrathin mos2 layer during electrocatalytic hydrogen
  evolution with a crystalline sno2 underlayer},}\ }\href {\doibase
  10.1039/D1RA00877C} {\bibfield  {journal} {\bibinfo  {journal} {RSC Adv.}\
  }\textbf {\bibinfo {volume} {11}},\ \bibinfo {pages} {17985--17992} (\bibinfo
  {year} {2021})}\BibitemShut {NoStop}%
\bibitem [{\citenamefont {Ruiz~M}\ \emph {et~al.}(2023)\citenamefont {Ruiz~M},
  \citenamefont {Terlinden}, \citenamefont {Engelhardt}, \citenamefont
  {Magnabosco}, \citenamefont {Papastavrou}, \citenamefont {Vogel},
  \citenamefont {Thommes},\ and\ \citenamefont {Bachmann}}]{Ruiz2023}%
  \BibitemOpen
  \bibfield  {author} {\bibinfo {author} {\bibfnamefont {C.~V.}\ \bibnamefont
  {Ruiz~M}}, \bibinfo {author} {\bibfnamefont {M.}~\bibnamefont {Terlinden}},
  \bibinfo {author} {\bibfnamefont {M.}~\bibnamefont {Engelhardt}}, \bibinfo
  {author} {\bibfnamefont {G.}~\bibnamefont {Magnabosco}}, \bibinfo {author}
  {\bibfnamefont {G.}~\bibnamefont {Papastavrou}}, \bibinfo {author}
  {\bibfnamefont {N.}~\bibnamefont {Vogel}}, \bibinfo {author} {\bibfnamefont
  {M.}~\bibnamefont {Thommes}}, \ and\ \bibinfo {author} {\bibfnamefont
  {J.}~\bibnamefont {Bachmann}},\ }\bibfield  {title} {\enquote {\bibinfo
  {title} {Preparation and surface functionalization of a tunable porous system
  featuring stacked spheres in cylindrical pores},}\ }\href {\doibase
  https://doi.org/10.1002/admi.202300436} {\bibfield  {journal} {\bibinfo
  {journal} {Advanced Materials Interfaces}\ }\textbf {\bibinfo {volume}
  {10}},\ \bibinfo {pages} {2300436} (\bibinfo {year} {2023})},\ \Eprint
  {http://arxiv.org/abs/https://onlinelibrary.wiley.com/doi/pdf/10.1002/admi.202300436}
  {https://onlinelibrary.wiley.com/doi/pdf/10.1002/admi.202300436} \BibitemShut
  {NoStop}%
\bibitem [{\citenamefont {Yarema}\ \emph {et~al.}(2014)\citenamefont {Yarema},
  \citenamefont {W{\"o}rle}, \citenamefont {Rossell}, \citenamefont {Erni},
  \citenamefont {Caputo}, \citenamefont {Protesescu}, \citenamefont {Kravchyk},
  \citenamefont {Dirin}, \citenamefont {Lienau}, \citenamefont {von Rohr},
  \citenamefont {Schilling}, \citenamefont {Nachtegaal},\ and\ \citenamefont
  {Kovalenko}}]{Yarema2014}%
  \BibitemOpen
  \bibfield  {author} {\bibinfo {author} {\bibfnamefont {M.}~\bibnamefont
  {Yarema}}, \bibinfo {author} {\bibfnamefont {M.}~\bibnamefont {W{\"o}rle}},
  \bibinfo {author} {\bibfnamefont {M.~D.}\ \bibnamefont {Rossell}}, \bibinfo
  {author} {\bibfnamefont {R.}~\bibnamefont {Erni}}, \bibinfo {author}
  {\bibfnamefont {R.}~\bibnamefont {Caputo}}, \bibinfo {author} {\bibfnamefont
  {L.}~\bibnamefont {Protesescu}}, \bibinfo {author} {\bibfnamefont {K.~V.}\
  \bibnamefont {Kravchyk}}, \bibinfo {author} {\bibfnamefont {D.~N.}\
  \bibnamefont {Dirin}}, \bibinfo {author} {\bibfnamefont {K.}~\bibnamefont
  {Lienau}}, \bibinfo {author} {\bibfnamefont {F.}~\bibnamefont {von Rohr}},
  \bibinfo {author} {\bibfnamefont {A.}~\bibnamefont {Schilling}}, \bibinfo
  {author} {\bibfnamefont {M.}~\bibnamefont {Nachtegaal}}, \ and\ \bibinfo
  {author} {\bibfnamefont {M.~V.}\ \bibnamefont {Kovalenko}},\ }\bibfield
  {title} {\enquote {\bibinfo {title} {Monodisperse colloidal gallium
  nanoparticles: Synthesis, low temperature crystallization, surface plasmon
  resonance and li-ion storage},}\ }\href {\doibase 10.1021/ja506712d}
  {\bibfield  {journal} {\bibinfo  {journal} {Journal of the American Chemical
  Society}\ }\textbf {\bibinfo {volume} {136}},\ \bibinfo {pages}
  {12422--12430} (\bibinfo {year} {2014})}\BibitemShut {NoStop}%
\bibitem [{\citenamefont {Rahim}\ \emph {et~al.}(2022)\citenamefont {Rahim},
  \citenamefont {Tang}, \citenamefont {Christofferson}, \citenamefont {Kumar},
  \citenamefont {Meftahi}, \citenamefont {Centurion}, \citenamefont {Cao},
  \citenamefont {Tang}, \citenamefont {Baharfar}, \citenamefont {Mayyas},
  \citenamefont {Allioux}, \citenamefont {Koshy}, \citenamefont {Daeneke},
  \citenamefont {McConville}, \citenamefont {Kaner}, \citenamefont {Russo},\
  and\ \citenamefont {Kalantar-Zadeh}}]{Rahim2022}%
  \BibitemOpen
  \bibfield  {author} {\bibinfo {author} {\bibfnamefont {M.~A.}\ \bibnamefont
  {Rahim}}, \bibinfo {author} {\bibfnamefont {J.}~\bibnamefont {Tang}},
  \bibinfo {author} {\bibfnamefont {A.~J.}\ \bibnamefont {Christofferson}},
  \bibinfo {author} {\bibfnamefont {P.~V.}\ \bibnamefont {Kumar}}, \bibinfo
  {author} {\bibfnamefont {N.}~\bibnamefont {Meftahi}}, \bibinfo {author}
  {\bibfnamefont {F.}~\bibnamefont {Centurion}}, \bibinfo {author}
  {\bibfnamefont {Z.}~\bibnamefont {Cao}}, \bibinfo {author} {\bibfnamefont
  {J.}~\bibnamefont {Tang}}, \bibinfo {author} {\bibfnamefont {M.}~\bibnamefont
  {Baharfar}}, \bibinfo {author} {\bibfnamefont {M.}~\bibnamefont {Mayyas}},
  \bibinfo {author} {\bibfnamefont {F.-M.}\ \bibnamefont {Allioux}}, \bibinfo
  {author} {\bibfnamefont {P.}~\bibnamefont {Koshy}}, \bibinfo {author}
  {\bibfnamefont {T.}~\bibnamefont {Daeneke}}, \bibinfo {author} {\bibfnamefont
  {C.~F.}\ \bibnamefont {McConville}}, \bibinfo {author} {\bibfnamefont
  {R.~B.}\ \bibnamefont {Kaner}}, \bibinfo {author} {\bibfnamefont {S.~P.}\
  \bibnamefont {Russo}}, \ and\ \bibinfo {author} {\bibfnamefont
  {K.}~\bibnamefont {Kalantar-Zadeh}},\ }\bibfield  {title} {\enquote {\bibinfo
  {title} {Low-temperature liquid platinum catalyst},}\ }\href {\doibase
  10.1038/s41557-022-00965-6} {\bibfield  {journal} {\bibinfo  {journal}
  {Nature Chemistry}\ }\textbf {\bibinfo {volume} {14}},\ \bibinfo {pages}
  {935--941} (\bibinfo {year} {2022})}\BibitemShut {NoStop}%
\bibitem [{\citenamefont {Arfken}\ and\ \citenamefont
  {Weber}(2005)}]{book_MathMeth}%
  \BibitemOpen
  \bibfield  {author} {\bibinfo {author} {\bibfnamefont {G.~B.}\ \bibnamefont
  {Arfken}}\ and\ \bibinfo {author} {\bibfnamefont {H.~J.}\ \bibnamefont
  {Weber}},\ }\href@noop {} {\emph {\bibinfo {title} {Atkins' Physical
  Chemistry}}}\ (\bibinfo  {publisher} {Elsevier Academic Press},\ \bibinfo
  {year} {2005})\BibitemShut {NoStop}%
\end{thebibliography}%

\end{document}